\documentclass[apj]{emulateapj}
\usepackage{natbib,graphicx,epstopdf,url}
\shortauthors{Plotkin et al.}

\begin{document}

\title{The X-ray Properties of Million Solar Mass Black Holes}
\shorttitle{10$^6$ $M_\odot$ Black Holes}

\author{
Richard.~M.~Plotkin\altaffilmark{1,2},
Elena~Gallo\altaffilmark{1},
Francesco Haardt\altaffilmark{3,4},
Brendan P.~Miller\altaffilmark{5},
Callum~J.~L.~Wood\altaffilmark{2},
Amy~E.~Reines\altaffilmark{1,6,8},
Jianfeng Wu\altaffilmark{1},
and Jenny E.~Greene\altaffilmark{7}
}

\altaffiltext{1}{Department of Astronomy, University of Michigan, 1085 South University Avenue, Ann Arbor, MI 48109, USA; richard.plotkin@curtin.edu.au}
\altaffiltext{2}{International Centre for Radio Astronomy Research (ICRAR), Curtin University, G.P.O. Box U1987, Perth, WA 6845, Australia}
\altaffiltext{3}{DiSAT, Universit\`{a} dell'Insubria, via Valleggio 11, I-22100 Como, Italy}
\altaffiltext{4}{INFN, Sezione di Milano--Bicocca, Piazza della Scienza 3, I-20126 Milano, Italy}
\altaffiltext{5}{Department of Chemistry and Physical Sciences, The College of St.\ Scholastica, Duluth, MN 55811, USA}
\altaffiltext{6}{National Optical Astronomy Observatory, 950 North Cherry Avenue, Tucson, AZ 85719, USA}
\altaffiltext{7}{Department of Astrophysics, Princeton University, Princeton, NJ 08544, USA}
\altaffiltext{8}{Hubble Fellow}

\newcommand{\nh}{N_{\rm H}}    
\newcommand{\lxledd}{L_X/L_{\rm Edd}}  
\newcommand{\lledd}{L_{\rm bol}/L_{\rm Edd}}   
\newcommand{\ledd}{L_{\rm Edd}}   
\newcommand{\lbol}{L_{\rm bol}}      
\newcommand{\ergs}{{\rm erg~s}^{-1}}
\newcommand{\rgas}{r_{\rm gas}}       

\newcommand{\mdot}{\dot{m}}
\newcommand{\Mdot}{\dot{M}}
\newcommand{\msun}{M_{\odot}}
\newcommand{\mstar}{M_{\star}}
\newcommand{\mbh}{M_{\rm BH}}

\newcommand{\lx}{L_{\rm X}}     
\newcommand{\lkev}{l_{\rm 2keV}}    
\newcommand{\luv}{l_{\rm 2500}}     

\newcommand{\smbh}{SMBH}   
\newcommand{\imbh}{mBH}      
\newcommand{\imbhagn}{mBH AGN}   
\newcommand{\bh}{black hole}            
\newcommand{\Bh}{Black hole}            

\newcommand{\ntarg}{seven}   
\newcommand{\ntargdig}{7}

\newcommand{\parent}{high-$\lledd$}    

\newcommand{\srconefull}{SDSS J003552.26$+$011249.4}
\newcommand{\srctwofull}{SDSS J081825.15$+$472950.3}
\newcommand{\srcthreefull}{SDSS J023310.79$-$074813.3}
\newcommand{\srcfourfull}{SDSS J083803.67$+$540642.2}
\newcommand{\srcfivefull}{SDSS J084013.23$+$412357.0}
\newcommand{\srcsixfull}{SDSS  J095306.81$+$365028.0}
\newcommand{\srcsevenfull}{SDSS J111547.46$+$502405.6}
\newcommand{\srconeshrt}{SDSS J0035}
\newcommand{\srctwoshrt}{SDSS J0818}
\newcommand{\srcthreeshrt}{SDSS J0233}
\newcommand{\srcfourshrt}{SDSS J0838}
\newcommand{\srcfiveshrt}{SDSS J0840}
\newcommand{\srcsixshrt}{SDSS  J0953}
\newcommand{\srcsevenshrt}{SDSS J1115}
\newcommand{\srconeobs}{15015}
\newcommand{\srctwoobs}{15016}
\newcommand{\srcthreeobs}{15017}
\newcommand{\srcfourobs}{15018}
\newcommand{\srcfiveobs}{15019}
\newcommand{\srcsixobs}{15020}
\newcommand{\srcsevenobs}{15021}
\newcommand{\srcone}{galaxy 7}
\newcommand{\srctwo}{galaxy 40}
\newcommand{\srcthree}{galaxy 19}
\newcommand{\srcfour}{galaxy 53}
\newcommand{\srcfive}{galaxy 56}
\newcommand{\srcsix}{galaxy 82}
\newcommand{\srcseven}{galaxy 111}
\newcommand{\srconedig}{7}
\newcommand{\srctwodig}{40}
\newcommand{\srcthreedig}{19}
\newcommand{\srcfourdig}{53}
\newcommand{\srcfivedig}{56}
\newcommand{\srcsixdig}{82}
\newcommand{\srcsevendig}{111}

\newcommand{\nar}{NewAR}
\defcitealias{greene07b}{GH07}
\defcitealias{dong12}{D12}        

\begin{abstract}
We present new  \textit{Chandra} X-ray observations of seven low-mass black holes ($\mbh \approx 10^6~\msun$) accreting at low-bolometric Eddington ratios between  $-2.0 \lesssim \log \lledd \lesssim -1.5$.   We compare the X-ray properties of these seven  low-mass active galactic nuclei (AGN) to a total of 73 other low-mass AGN in the literature with published \textit{Chandra} observations (with Eddington ratios extending from  $-2.0 \lesssim \log \lledd \lesssim -0.1$).  We do not find any statistical differences between  low- and high-Eddington ratio low-mass AGN in the distributions of their X-ray to ultraviolet  luminosity ratios ($\alpha_{\rm ox}$), or in their X-ray spectral shapes.  Furthermore, the  $\alpha_{\rm ox}$ distribution of low-$\lledd$ AGN displays an X-ray weak tail  that is also observed within  high-$\lledd$ objects.    Our results indicate that between  $-2 \lesssim \log \lledd \lesssim -0.1$, there is no systematic change in the structure of the accretion flow  for active galaxies hosting $10^6~\msun$ black holes.    We examine the accuracy of current bolometric luminosity estimates  for our  low-$\lledd$ objects with new \textit{Chandra} observations, and it is plausible that their Eddington ratios could be underestimated by up to an order of magnitude.  If so, then in analogy with weak emission line quasars, we suggest that accretion from a geometrically thick, radiatively inefficient `slim disk' could  explain their diverse properties in $\alpha_{\rm ox}$.  Alternatively,  if current Eddington ratios are in fact correct (or overestimated), then the X-ray weak tail would imply that there is diversity in disk/corona couplings among individual low-mass objects. Finally, we conclude by noting that the $\alpha_{\rm ox}$ distribution for low-mass black holes may have favorable consequences for the epoch of cosmic reionization being driven by AGN. \end{abstract}


\section{Introduction} 
\label{sec:intro}

Virtually every large galaxy  harbors a supermassive black hole (\smbh; $\mbh \approx 10^6-10^9~\msun$) in its nucleus \citep[e.g.,][]{kormendy13}.  However, understanding in detail the mechanism(s) that  formed the first primordial `seed' black holes, and determining the evolutionary path that those seeds took to grow to SMBH sizes, remains a fundamental problem \citep[e.g.,][]{greene12, natarajan14}.  We currently know of  $\sim$40 quasars at $z\gtrsim6$ with $>$10$^9~\msun$ SMBHs \citep[see, e.g.,][]{willott03, fan06, wu15}.   \Bh\ seeds produced by the collapse of Population III stars would have  $\mbh \lesssim$1000~$\msun$ \citep[e.g.,][]{madau01, bromm02}, and one generally needs to invoke super-Eddington accretion onto such low-mass seed \bh s in order to explain the presence of high-redshift quasars \citep{madau14}.   On the other hand, more massive seeds ($10^5-10^6~\msun$), as could be produced from the direct collapse of gas clouds at the centers of galaxies \citep[e.g.,][]{begelman06, lodato06},  provide another feasible way to cultivate    rapid  growth.  We refer to $10^5-10^6~\msun$ \bh s here as ``massive black holes'' (\imbh s).  

\begin{deluxetable*}{l c c c c c  c c c}
\tablecaption{Galaxy Properties and X-ray Observing Log}
\tablecolumns{8}
\renewcommand\arraystretch{1.2}
\tablehead{
                \colhead{Galaxy Name}      		&  
		\colhead{GH07 ID}  				&  
		\colhead{z} 					& 
		\colhead{$\log N_{\rm H,gal}$}		& 
		\colhead{$\log \luv$}                                  & 
		\colhead{$\log \mbh$}			& 
		\colhead{$\log \left(L_{\rm bol}/L_{\rm Edd}\right)$}			& 
		\colhead{ObsID}				& 
		\colhead{$\tau_{\rm exp}$}		\\  
		\colhead{(SDSS J)}                            & 
		\colhead{}         					& 
		\colhead{}						& 
		\colhead{(cm$^{-2}$)}				& 
		\colhead{erg s$^{-1}$ Hz$^{-1}$}        & 
		\colhead{$(\msun)$}				& 
		\colhead{}						& 
		\colhead{}						& 
		\colhead{(ks)}					 \\ 
		\colhead{(1)}					& 
		\colhead{(2)}					& 
		\colhead{(3)}					& 
		\colhead{(4)}					& 
		\colhead{(5)}					& 
		\colhead{(6)}					& 
		\colhead{(7)}					& 
		\colhead{(8)}					 & 
		\colhead{(9)}                         
}
\startdata
    003552.26$+$011249.4 &            7 &       0.0414 &        20.4 &          26.3         & 5.8 &       $-1.7$ &        15015 &        15.7 \\
    023310.79$-$074813.3 &           19 &       0.0310 &        20.5 &          26.7         & 6.1 &       $-1.6$ &        15017 &         10.0 \\
    081825.15$+$472950.3 &           40 &       0.0537 &        20.6 &          26.7         & 6.0 &       $-1.5$ &        15016 &        25.5 \\
    083803.67$+$540642.2 &           53 &       0.0295 &        20.6 &          26.6          &   6.2 &       $-1.8$ &        15018 &         9.0 \\
    084013.23$+$412357.0 &           56 &       0.0290 &        20.3 &          26.4          &  6.2 &       $-2.0$ &        15019 &         9.0 \\
    095306.81$+$365028.0 &           82 &       0.0491 &        20.1 &          26.6          &  6.0 &       $-1.6$ &        15020 &        20.5 \\
    111547.46$+$502405.6 &          111 &       0.0473 &        20.1 &          26.8         & 6.1 &       $-1.5$ &        15021 &        21.8 
\enddata
\tablecomments{
Column (1) SDSS galaxy designation.
Column (2) galaxy ID from \citetalias{greene07b}.
Column (3) spectroscopic redshift from the SDSS.
Column (4) log galactic column density along line of sight, from \citet{dickey90} \ion{H}{1} maps.
Column (5) log AGN continuum luminosity density at 2500~\AA\ (see Section~\ref{sec:aox}).
Column (6) log of the virial-scaled H$\alpha$-based black hole mass from \citetalias{greene07b}.
Column (7) log Eddington ratio from \citetalias{greene07b}.
Column (8) \textit{Chandra} Observation ID.
Column (9) effective exposure time of \textit{Chandra} observation.}
\label{tab:obslog}
\end{deluxetable*}
\renewcommand\arraystretch{1}

 Over the past 10-15 years, samples of \imbh-powered active galactic nuclei (AGN) have rapidly evolved  from discoveries of a handful of individual objects --- such as NGC 4395 \citep{filippenko03}, POX 52 \citep{barth04}, and Henize 2-10 \citep{reines11} --- to the production of systematically assembled catalogs drawn from large sky surveys \citep[e.g.,][]{greene07b, barth08, dong12, reines13, schramm13, moran14}.   The largest \imbh\ catalogs so far have utilized the Sloan Digital Sky Survey \citep[SDSS;][]{york00}, particularly \citet[][hereafter \citetalias{greene07b}]{greene07b} and \citet{dong12} who select \imbh\ AGN based on the  presence of broad H$\alpha$ emission in  optical spectra.  They  use  the widths and luminosities of broad H$\alpha$  to estimate \bh\ masses and bolometric luminosities via virial scaling techniques \citep{greene05}.  Both studies require $\mbh < 2\times10^6~\msun$, and their catalogs include 229 \citepalias{greene07b} and 309 \citep{dong12} objects.   \citet{reines13} also recover a sizable number of \imbh\ candidates (151 objects) with a slightly different approach.  They specifically target dwarf galaxies by  restricting host galaxy stellar masses to $\mstar<3\times10^9 \msun$.  The vast majority (136) of their \imbh\ AGN candidates are then selected via photoionization signatures of black hole activity via narrow emission line ratios (only a small fraction of their 151 AGN candidates display broad H$\alpha$).  The smallest \imbh\ discovered so-far through optical searches weighs in at only $5\times10^4~\msun$ \citep{reines13, baldassare15}.  
 
 It is natural to turn to the X-ray waveband for insight into the properties of the accretion flows that feed \imbh s, since X-ray emission is a nearly universal feature of accretion.  So far, the most extensive X-ray follow-up is based on 67  objects that were observed  with the \textit{Chandra} X-ray telescope (\citealt{greene07, desroches09, dong12a}, hereafter \citetalias{dong12}; also see \citealt{ dewangan08, miniutti09} for an \textit{XMM-Newton} view of about a dozen \imbh-powered AGN).   These X-ray observations show that \citetalias{greene07b} \imbh s tend to have higher X-ray to ultraviolet (UV) luminosity ratios on average compared to ($10^8-10^9~\msun$) Type 1 quasars.  The higher X-ray to UV ratios are generally consistent with expectations from accretion disk/corona models \citep[e.g.,][]{haardt93, done12}, where  lower black hole masses yield higher accretion disk temperatures, which results in less UV emission near 2500~\AA\ (combined with less efficiently Compton-cooled coronae, this yields higher X-ray to UV flux ratios; see \citetalias{dong12}).   However,   as a population, \imbh\ AGN show a  large dispersion in their X-ray to UV luminosity ratios, and they display a  puzzling  ``X-ray weak'' tail \citepalias{dong12}.  
  
  The above X-ray studies only targeted \imbh s at high Eddington ratios\footnote{$L_{\rm bol}$ is the bolometric luminosity, which \citetalias{greene07b} calculate based on the luminosity of broad H$\alpha$ emission (see their Section~3).  The Eddington luminosity, $\ledd = 1.26\times10^{38}\left(M/\msun\right)~\ergs$ for ionized Hydrogen, is the maximum luminosity before radiation pressures halts accretion, assuming a spherical geometry and isotropic radiation.} 
  ($\log \lledd \gtrsim -1$).   Here, we present new X-ray observations of seven \citetalias{greene07b} \imbh s at Eddington ratios up to an order of magnitude lower ($-2.0\lesssim \log \lledd \lesssim -1.5$).   By increasing the dynamic range in Eddington ratio, we can search for trends between X-ray properties and $\lledd$.    We describe our sample selection and X-ray observations/analysis in Section~\ref{sec:sample}.   X-ray results are presented in Section~\ref{sec:res}, where we also  discuss the optical emission line properties of our targets  (Section~\ref{sec:optical}), as well as bolometric corrections for \imbh s (Section~\ref{sec:lbol}).  Our results are discussed in Section~\ref{sec:disc}.   For consistency with previous \textit{Chandra} follow-up, all optical spectroscopic measurements are taken from \citetalias{greene07b}, and we generally follow the recipes of \citetalias{dong12} when deriving X-ray  properties.  We adopt the same cosmology as \citetalias{greene07b} and \citetalias{dong12}:  $H_0=71$ km s$^{-1}$ Mpc$^{-1}$, $\Omega_m=0.27$, and $\Omega_\Lambda=0.75$ \citep{spergel03}.  
   
 \section{Sample and Observations}
 \label{sec:sample}
 Chandra targets were selected from the \citetalias{greene07b} catalog of \imbh s, from which we consider low-Eddington ratio AGN  with $\log \lledd < -1.5$ (for estimating bolometric luminosities, \citetalias{greene07b} assume that $L_{\rm bol} = 9.8 L_{5100}$, where $L_{5100}$ is the AGN luminosity at 5100~\AA; please refer to Sections \ref{sec:aox} and \ref{sec:lbol} for details).  There are 17 \imbh s in \citetalias{greene07b} with such low Eddington ratios.  To improve X-ray efficiency, we further restrict our target list to only include nearby galaxies ($z<0.055$, which is comparable to the average  redshift of \citetalias{greene07b} objects targeted by \citetalias{dong12}), providing seven \textit{Chandra} targets.   These seven galaxies were observed by \textit{Chandra} during cycle-14 (proposal ID 14700673; PI Gallo).  The galaxy properties and observations are summarized in Table~\ref{tab:obslog}.  Throughout the text, we refer to each galaxy by the catalog number assigned by \citetalias{greene07b} (see Column 2 of Table~\ref{tab:obslog}).  
 
\begin{figure*}
\begin{center}
\includegraphics[scale=0.8]{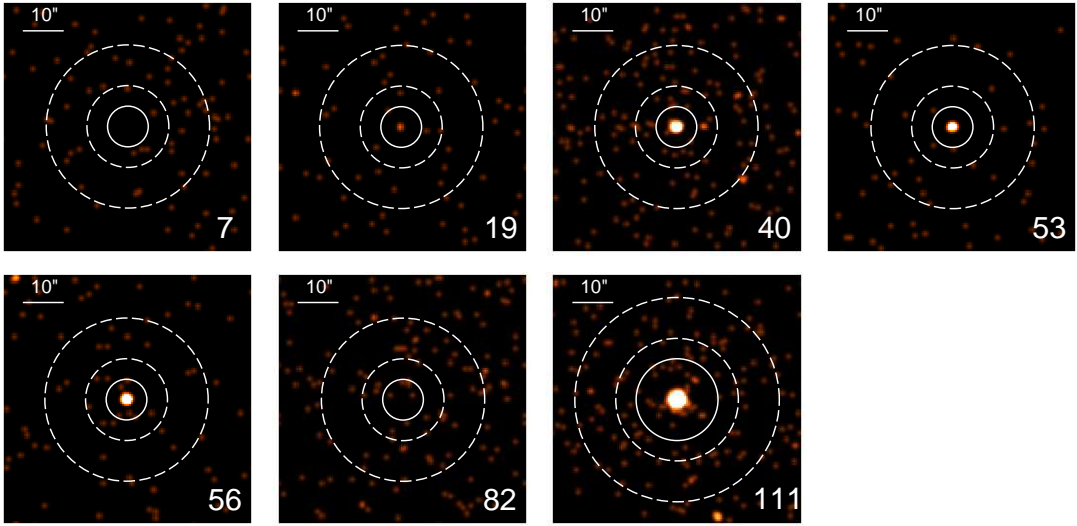}
\caption{Chandra images of each galaxy.  Each image is $1 \times 1 \arcmin$ on a side, smoothed using a Gaussian kernel with $\sigma=3$ pixels.  The regions for measuring source and background counts are shown as white solid circles and dashed annuli, respectively. An X-ray source is detected at the $>$95\% confidence level within galaxies \srctwodig, \srcfourdig, \srcfivedig, and \srcsevendig.}
\label{fig:xrayims}
\end{center}
\end{figure*}

\begin{deluxetable*}{l c c c c c c c c}
\tablecaption{X-ray Properties}
\tablecolumns{9}
\tablewidth{\linewidth}
\renewcommand\arraystretch{1.2}

\tablehead{
                \colhead{GH07 ID}       	  		&  
                \colhead{$N_s$}                         & 
                \colhead{$N_h$}                         & 
		\colhead{$R_s$}  				&  
		\colhead{$R_h$} 				& 
		\colhead{$\log f_{s,u}$} 			& 
		\colhead{$\log f_{h,u}$}			& 
		\colhead{$\log L_s$}				& 
		\colhead{$\log L_h$}				\\ 
                \colhead{}       	  		&  
                \colhead{(counts)}                 & 
                \colhead{(counts)}                 & 
		\colhead{(counts ks$^{-1}$)}  		&  
		\colhead{(counts ks$^{-1}$)} 		& 
		\colhead{(erg s$^{-1}$ cm$^{-2}$)} 	& 
		\colhead{(erg s$^{-1}$ cm$^{-2}$)}	& 
		\colhead{(erg s$^{-1}$)}			& 
		\colhead{(erg s$^{-1}$)}			\\ 
		\colhead{(1)}					& 
		\colhead{(2)}					& 
		\colhead{(3)}					& 
		\colhead{(4)}					& 
		\colhead{(5)}					& 
		\colhead{(6)}					& 
		\colhead{(7)}					& 
		\colhead{(8)}					 & 
		\colhead{(9)}					 
}
\startdata
           7 &                  \nodata &                  \nodata &                $<$$0.2$ &                $<$$0.3$ &              $<$$-14.9$ &              $<$$-14.3$ &                $<$$39.7$ &                $<$$40.3$ \\
          19 &                  \nodata &                  \nodata &                $<$$0.3$ &                $<$$0.4$ &              $<$$-14.8$ &              $<$$-14.2$ &                $<$$39.5$ &                $<$$40.2$ \\
          40 &        $54.7 \pm 12.7$ &         $25.2 \pm 8.9$ &          $2.2 \pm 0.5$ &          $1.0\pm 0.4$ &        $-14.0 \pm 0.1$ &        $-13.7 \pm 0.2$ &                     40.8 &                     41.1 \\
          53 &         $12.8 \pm 5.3$ &         $14.2 \pm 5.6$ &          $1.4 \pm 0.6$ &          $1.6 \pm 0.6$ &        $-13.4 \pm 0.2$ &        $-13.5 \pm 0.2$ &                     40.9 &                     40.8 \\
          56 &        $38.5 \pm 10.8$ &         $18.3 \pm 7.7$ &          $4.3 \pm 1.2$ &          $2.1 \pm 0.9$ &        $-13.7 \pm 0.1$ &        $-13.4 \pm 0.2$ &                     40.6 &                     40.9 \\
          82 &                  \nodata &                  \nodata &                $<$$0.2$ &                $<$$0.2$ &              $<$$-15.0$ &              $<$$-14.4$ &                $<$$39.7$ &                $<$$40.4$ \\
         111 &       $315.7 \pm 29.8$ &       $791.1 \pm 46.8$ &         $14.5 \pm 1.4$ &         $36.3 \pm 2.2$ &        $-12.3 \pm 0.04$ &        $-12.0 \pm 0.03$ &                     42.4 &                     42.8 
 \enddata
\tablecomments{
Column (1) \citetalias{greene07b} galaxy ID.
Column (2) net counts in the soft energy band (0.5-2~keV).  Error bars are at the 90\% confidence level, assuming Poisson statistics \citep{gehrels86}.  Blank values denote non-detections.
Column (3) net counts in the hard energy band (2-8~keV). 
Column (4) net count rate in the soft energy band (0.5-2~keV).  Upper limits are provided for non-detections (95\% level).
Column (5) net count rate in the hard energy band (2-8~keV).
Column (6) logarithm of unabsorbed soft X-ray flux (0.5-2~keV), assuming an absorbed power-law model.  If available, we adopt the best-fit column densities and photon indices in Table~\ref{tab:xrayspec}.  Otherwise, we use the Galactic column density in Table~\ref{tab:obslog} and $\Gamma=2$ (the average photon index for \citetalias{greene07b} AGN in \citetalias{dong12}).     Error bars are based on propagating the statistical errors on the net count rates, and they do not include errors on the emission model parameters.  All fluxes and luminosities for \srcseven\ are corrected for pileup. 
Column (7) logarithm of unabsorbed hard X-ray flux (2-8~keV).
Column (8) logarithm of unabsorbed luminosity in the soft band (0.5-2 keV).
Column (9) logarithm of unabsorbed luminosity in the hard band (2-8~keV).}
\label{tab:xraycnt}
\end{deluxetable*}

 \subsection{X-ray Analysis}
 \label{sec:obs}
 Each galaxy was placed at the aimpoint of the S3 chip on the Advanced CCD Imaging Spectrometer \citep[ACIS;][]{garmire03}.  The data were telemetered in {\tt FAINT} mode and    reduced with the \textit{Chandra} Interactive Analysis of Observations software v4.7 ({\tt CIAO}; \citealt{fruscione06}).  We  reprocessed each \textit{Chandra} observation, applying the latest calibration files (CALDB v4.6.7).  The event files were then filtered to only retain events with grades 0,2,3,4,6 and energies between 0.3--8~keV.  We examined light curves of the ACIS background, and we removed time periods where the background deviated by more than 3$\sigma$ from the mean, using the {\tt lc\_sigma\_clip} routine in {\tt CIAO}.  Only 200 s were removed from ObsIDs 15015 and 15016 (galaxies \srconedig\ and \srctwodig), and 400 s from ObsID 15020 (\srcsix).  The effective exposure times for each observation are listed in Table~\ref{tab:obslog}.
 
We ran the automated point-source detection tool {\tt wavdetect} on  0.3-8~keV images of the S3 chip, adopting  wavelet scales of 1.0, 1.4, 2.0, 2.8, and 4.0 pixels, a point spread function (psf) map calculated at 2.3~keV, and a significance threshold of 10$^{-6}$ (which corresponds to one expected false detection across the S3 chip).   Within the positional accuracy of a source located at the aimpoint of the ACIS detector ($\sim$0$\farcs$4), four observations contained an X-ray point source at a location consistent with the position of the SDSS optical galaxy (galaxies \srctwodig, \srcfourdig, \srcfivedig, and \srcsevendig).  Images of each \textit{Chandra} target (over 0.5-8 keV) are displayed in Figure~\ref{fig:xrayims}.  All X-ray sources are point-like, with no signs of nearby diffuse emission or nearby (off-nuclear) neighbors.  We associate all four X-ray detections with the AGN (see Sections \ref{sec:res}).

Next, we created soft- (0.5-2~keV) and hard-band (2-8~keV) images of the S3 chip (these bands were chosen to match the analysis of \citetalias{dong12}).   The numbers of soft and hard X-ray counts were extracted over circular regions with 5$\arcsec$ radii, centered on the nuclear X-ray source (or the SDSS optical position for the three observations where {\tt wavdetect} did not find an X-ray source).  The local background was estimated over an annulus concentric with the source extraction region, with inner and outer  radii of 10 and 20$\arcsec$, respectively.  The X-ray source in \srcseven\ is significantly brighter than the others, so we adopted a 10$\arcsec$ radius source extraction circle, and a background annulus with inner and outer radii 15 and 25$\arcsec$ for that observation.  In Table~\ref{tab:xraycnt}, we list the total numbers of counts extracted within each circular region, and the net count rates in both the soft and hard bands.  We typically obtained $\approx$10--50 net counts in each band, except for \srcseven\ where we obtained $\approx$300 and 800 net soft- and hard-counts, respectively.

 For the four galaxies with associated X-ray emission, we assign uncertainties on the total numbers of counts using Poisson statistics \citep{gehrels86}.   For the other three galaxies, we use the estimated number of background counts in each source aperture (scaled from the measured number of background counts in each sky annulus) and the Bayesian method of \citet{kraft91} to calculate the number of counts required to detect an X-ray source at the 95\% confidence level (typically 3-6 counts).  We confirm that no X-ray point source is present for these three galaxies, and we include 95\% upper limits on the net count rates in Table~\ref{tab:xraycnt}.
 
 Finally, we estimate unabsorbed X-ray fluxes (and luminosities)  using the Portable, Interactive Multi-Mission Simulator (PIMMS)\footnote{\url{http://heasarc.gsfc.nasa.gov/docs/software/tools/pimms.html}} and  a power-law model, where we adopt  the best-fit photon index ($\Gamma$) and column density ($\nh$) described in Section~\ref{sec:xrayspec}.  For the three galaxies lacking an X-ray source, we provide flux and luminosity upper limits by assuming $\Gamma=2.0$ (the average value from \citetalias{dong12}) and the Galactic $\nh$\ from the \citet{dickey90} \ion{H}{1} maps (see Table~\ref{tab:obslog}).

 \subsection{X-ray Spectral Analysis}
 \label{sec:xrayspec}
For the four galaxies with nuclear X-ray detections, we extract  X-ray spectra over  0.3-8~keV  using the {\tt CIAO} tool {\tt specextract}.   We use the Interactive Spectral Interpretation System \citep[ISIS;][]{houck00} to analyze the spectra.  Given the typically low number of counts, and in order to compare to the spectral analysis of \citetalias{dong12}, we only attempt to fit an absorbed power-law model to each spectrum ({\tt phabs*powerlaw} in ISIS).  We utilize Cash statistics \citep{cash79}, with the background included in each spectral fit.

\begin{figure}
\begin{center}
\includegraphics[scale=0.55]{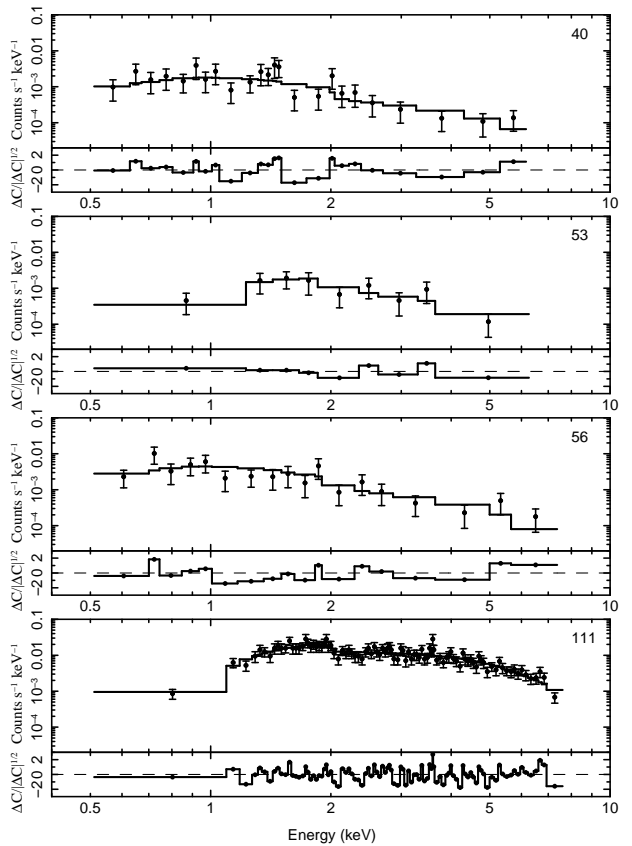}
\caption{Spectral fits ({\tt phabs*powerlaw}) to the four \textit{Chandra} observations with X-ray detections (with residuals displayed as $\Delta C / \left|\Delta C\right|^{0.5}$, where $C$ is the Cash statistic).  The best-fit parameters are listed in Table~\ref{tab:xrayspec}.}
\label{fig:xspec}
\end{center}
\end{figure}

The spectral fits are shown in Figure~\ref{fig:xspec}, and the best-fit spectral parameters are reported in Table~\ref{tab:xrayspec}, including the photon index ($\Gamma$) and column density ($\nh$).   Error bars are quoted at the 90\% level (corresponding to a change in the Cash statistic of $\Delta C=2.71$ for one parameter of interest).  Two sources (galaxies  \srcfourdig\ and \srcsevendig) display moderate absorption, with intrinsic column densities $\approx$10$^{22}$ cm$^{-2}$.  For the other two galaxies with X-ray detections (\srctwodig\ and \srcfivedig), the column density converges toward zero during the spectral fits.  If we instead fix $\nh$\ to the Galactic value from the \citet{dickey90} maps for those two sources, the best-fit $\Gamma$ values do not change significantly for either source ($\Gamma=1.63_{-0.25}^{+0.50}$ and $1.63_{-0.47}^{+0.33}$ for galaxies \srctwodig\ and \srcfivedig, respectively).  For consistency, in Table~\ref{tab:xrayspec} we   always quote $\Gamma$ values for the fits where $\nh$\ is allowed to vary.   

The X-ray source in \srcseven\ has a high count rate, and it appears to  suffer mildly from the effects of photon pileup.  For sources with high count rates, two or more photons may hit a CCD detector region before the frame is read out (every 3.2 sec for our observations), and the multiple photons are  registered as only a single event.  One effect of pileup is that the observed X-ray spectrum may appear harder than the intrinsic one, because of energy migration, where the registered event has an energy equal to the sum of the multiple  ``piled'' photons.  To correct for this effect, the best-fit model parameters for \srcseven\ in Table~\ref{tab:xrayspec} are reported using the \citet{davis01} pileup model.\footnote{For this observation, we use a spectrum extracted from an event file that includes all events with energies $>$0.3~keV (i.e., we do not apply an 8~keV high-energy filter); see \url{http://cxc.harvard.edu/ciao/why/filter\_energy.html}.} 
 The pileup fraction is  $f_{\rm pile}=0.05$ (as determined from the best-fit pileup model).   The numbers of net counts and count rates reported for \srcseven\ in Table~\ref{tab:xraycnt} are the observed numbers, without any pileup corrections; the reported fluxes and luminosities are pileup corrected  (calculated within {\tt ISIS} using the best-fit pileup model).
 
\begin{deluxetable}{l c c c c}
\tablecaption{Broadband and X-ray Spectral Properties}
\tablecolumns{5}
\renewcommand\arraystretch{1.2}

\tablehead{
                \colhead{GH07 ID}       	  		&  
		\colhead{$\log N_{\rm H,fit}$}		& 
		\colhead{$\Gamma_{\rm fit}$}		&  
		\colhead{$\alpha_{\rm ox}$}		& 
		\colhead{$\Delta \alpha_{\rm ox}(\sigma)$}  \\ 
                \colhead{}       	  		&  
		\colhead{(cm$^{-2}$)}		 	& 
		\colhead{}						&  
		\colhead{}						& 
		\colhead{}  					\\  
		\colhead{(1)}					& 
		\colhead{(2)}					& 
		\colhead{(3)}					& 
		\colhead{(4)}					& 
		\colhead{(5)}					
}
\startdata
            7 &                  \nodata &                  \nodata  & $<$$       -1.7$ & $<$$       -0.7(-8.0)$ \\
          19 &                  \nodata &                  \nodata  & $<$$       -1.9$ & $<$$       -0.9(-9.4)$ \\
          40 &                  \nodata &          $1.5 \pm 0.2$  &    $       -1.4$ &    $       -0.3(-3.6)$ \\
          53 &         $22.1 \pm 0.4$ &          $2.3 \pm 1.5$  &    $       -1.4$ &    $       -0.4(-3.9)$ \\
          56 &                  \nodata &          $1.6 \pm 0.4$  &    $       -1.4$ &    $       -0.4(-4.0)$ \\
          82 &                  \nodata &                  \nodata  & $<$$       -1.8$ & $<$$       -0.8(-8.5)$ \\
         111\footnote{Spectral fit performed with the \citet{davis01} pileup model, with a grade migration parameter $\alpha=0.6$.  The probability of retaining $n$ events that are `piled' together is $p\sim \alpha^{n-1}$.} &         $22.2 \pm 0.1$ &          $1.5 \pm 0.3$  &    $       -0.8$ &    $        0.3(2.9)$ 
 \enddata
\tablecomments{
Column (1) \citetalias{greene07b} galaxy ID.
Column (2) logarithm of the best-fit column density.  Blank values indicate that no spectral fit was performed (IDs 7, 19, and 82), or that the best-fit column density converged toward zero (IDs 40 and 56).
Column (3) best-fit photon index, $N_E = N_0\left(E/E_0\right)^{-\Gamma}$.  Blank values indicate that no spectral fit was performed.
Column (4) X-ray to UV luminosity ratio $\alpha_{\rm ox} = -0.3838 \log \left(\luv/\lkev\right)$,  where $\lkev$ and $\luv$ are (unabsorbed) X-ray and UV luminosity densities at 2~keV and 2500~\AA, respectively (see Section~\ref{sec:aox}).
Column (5) the difference between $\alpha_{\rm ox}$ and the value expected from the \citet{just07} $\alpha_{\rm ox}-\luv$ relation.  The statistical significance in parentheses is based on the rms scatter of $\alpha_{\rm ox}$ as a function of $\luv$, as presented in Table~5 of \citet{steffen06}.  More negative numbers are ``X-ray weaker.'' }
\label{tab:xrayspec}
\end{deluxetable}

\subsection{Other \imbh s with Chandra Coverage}
\label{sec:samp:parent}

Throughout the remainder of the text, we compare our observations to  a total of 73 other \imbh s with \textit{Chandra} coverage.  The bulk of this comparison sample includes 67 higher Eddington ratio  objects, composed of  49 \textit{Chandra} observations originally published by  \citetalias{dong12}, 10 from \citealt{greene07}, and 8 from  \citealt{desroches09}.  (We refer to these as \citetalias{greene07b} AGN, even though 18/67 of these objects were originally identified as AGN by \citealt{greene04}).  We collectively refer to these 67 objects as the \textit{\parent\ sample}.   X-ray information for these 67 objects are taken directly from \citetalias{dong12} or \citet{desroches09}, unless stated otherwise.

The remaining 6 archival objects come from a search  of the literature for lower-Eddington ratio \imbh s with \textit{Chandra} coverage.   Observations of four objects are presented by \citet{yuan14}, whose sample includes two sources selected from the \citet{dong12} catalog (SDSS~J004042.10$-$110957.6 and SDSS~J112637.74+513423.0; both of these sources also appear in  \citetalias{greene07b}), and two sources that were selected by \citet{yuan14} from the SDSS Data Release 5\footnote{Both the \citetalias{greene07b} and \citet{dong12} samples were based on SDSS Data Release 4.  \citet{yuan14} identified these two sources by applying similar selection algorithms as \citet{dong12}.} (SDSS J074345.47+480813.5 and SDSS J130456.95+395529.7).  These four sources span $-2.0 < \lledd < -1.3$, as calculated by \citet[][see their Table 1]{yuan14}.  We also consider \citetalias{greene07b} objects with new \textit{Chandra} observations presented by \citet{gultekin14}.  To avoid duplicating a similar parameter space as \citetalias{dong12}, we exclude  objects from \citet{gultekin14} with $\log \lledd > -1$.   Of their six sources with new \textit{Chandra} observations, two  remain --- SDSS~J121629.13+601823.5 ($\log \lledd = -1.3$) and SDSS~J132428.24+044629 ($\log \lledd = -1.4$).  Although both of these sources were initially identified as \imbh\ AGN by \citetalias{greene07b}, neither would have been included in our cycle-14 \textit{Chandra} program because their Eddington ratios are above our $\log \lledd < -1.5$ criterion (plus, SDSS J121629.13+601823.5 is at too high of a redshift; $z=0.0601$).  We include them here nevertheless to improve statistics, as they span a similar range in $\log \lledd$ as the four  objects from \citet{yuan14}.   X-ray information for these combined six archival observations are taken directly from \citet{yuan14} and \citet{gultekin14}, unless stated otherwise.

\subsection{X-ray Non-Detections and Stacking Analysis}
\label{sec:xraystack}

Three of our \textit{Chandra} targets do not have X-ray detections, and a total of 15 other objects compiled in our comparison sample (Section~\ref{sec:samp:parent}) are reported as non-detections in the literature (12 from the \parent\ sample, two from \citealt{yuan14}, and one from \citealt{gultekin14}).  Since different detection thresholds are applied in each paper, we re-analyze all 15 archival observations here, so that we can compare upper limits on X-ray fluxes consistently between our analysis and the archival ones.

We repeat a similar data reduction on these 15 archival objects as described in Section~\ref{sec:obs}, with the primary difference being that we use a $1\farcs5$ aperture for extracting source counts, in order to minimize the background in each aperture.  Results of the photometry in soft (0.5-2~keV) and hard (2-8~keV) images are presented in Table~\ref{tab:xlimphot}, where we also repeat the analysis for our three Cycle 15 \textit{Chandra} targets adopting $1\farcs5$ apertures.  In the final column of Table~\ref{tab:xlimphot}, we tabulate whether each source is considered an X-ray detection at the 95\% and 99\% level  over the full 0.5-8~keV band (according to the confidence interval tables in \citealt{kraft91}).    We find 6/18 objects are detected at the 95\% level, and 4/18 at the 99\% level in each 0.5-8~keV image.\footnote{Our X-ray photometry on these 15 archival observations are consistent with that already reported in the literature.  Differences in which ones are considered detections are a result of varying definitions of detection thresholds.} 

For the six objects detected at the 95\% confidence level, we adopt the fluxes quoted in Table~\ref{tab:xlimphot} throughout the remainder of the text.  The X-ray fluxes are calculated within PIMMS, assuming an absorbed power-law with $\Gamma=2.0$ and $\nh$ set to the Galactic value from \citet{dickey90}.  The fluxes are calculated over the full 0.5-8~keV band, after applying a 92\% aperture correction, in order to account for the fraction of X-ray photons excluded by our choice of a $1\farcs5$ source aperture (these aperture corrections are based on the enclosed energy fraction at 2.3~keV at the  \textit{Chandra} ACIS-S3 aimpoint; including these aperture corrections produces fluxes consistent with the larger apertures that we adopted in Section~\ref{sec:obs}).  For the 12 sources that remain undetected at the 95\% confidence level, we include 95\% upper limits on the full-band X-ray fluxes in Table~\ref{tab:xlimphot}.

\begin{deluxetable*}{l c c c c c c c c c c}
\tablecaption{Photometry of Sources Reported as X-ray Non-detections in the Literature}
\tablecolumns{11}
\renewcommand\arraystretch{1.2}
\tablehead{
		\multicolumn{3}{c}{}    &
		\multicolumn{2}{c}{Soft (0.5-2 keV)}  &
		\multicolumn{2}{c}{Hard (2-8 keV)}   &
		\colhead{Full (0.5-8 keV)}            & 
		\multicolumn{2}{c}{Detection?} & 
		\colhead{} \vspace{0.2cm} \\
                 \colhead{Galaxy Name}       	  			&  
		\colhead{ObsID}				& 
		\colhead{$\tau_{\rm exp}$}		&  
		\colhead{$C_{s,{\rm tot}}$}		& 
		\colhead{$B_s$}                		& 
		\colhead{$C_{h,{\rm tot}}$}                 & 
		\colhead{$B_h$}                                 & 
		\colhead{$\log f_{0.5-8}$}                     & 
		\colhead{95\%}                                    & 
		\colhead{99\%}                                    & 
		\colhead{Ref}                                       \\ 
                \colhead{(SDSS J)}       	  		&  
		\colhead{}					 	& 
		\colhead{(ks)}					&  
		\colhead{(counts)}				& 
		\colhead{(counts)}  				& 
		\colhead{(counts)}  				& 
		\colhead{(counts)}  				& 
		\colhead{(erg s$^{-1}$ cm$^{-2}$)}	& 
		\colhead{}						& 
		\colhead{}						& 
		\colhead{}						\\ 
		\colhead{(1)}					& 
		\colhead{(2)}					& 
		\colhead{(3)}					& 
		\colhead{(4)}					& 
		\colhead{(5)}					& 
		\colhead{(6)}					& 
		\colhead{(7)}					& 
		\colhead{(8)}					& 
		\colhead{(9)}					& 
		\colhead{(10)}					& 
		\colhead{(11)}					 
}
\startdata
003552.26+011249.4	&	15015	&	15.9	&	0	&	0.06	&	0	&	0.09	&	$< -14.8$	&	N	&	N	&	this work	\\
004042.10$-$110957.6	&	9235	&	4.7	&	1	&	0.02	&	1	&	0.1	&	$< -14.1$	&	N	&	N	&	Y14	\\
023310.79$-$074813.3	&	15017	&	10.0	&	0	&	0.06	&	0	&	0.06	&	$< -14.6$	&	N	&	N	&	this work	\\
094310.12+604559.1	&	5661	&	5.0	&	2	&	0.07	&	0	&	0.1	&	$< -14.0$	&	N	&	N	&	GH07c	\\
095306.81+365028.0	&	15020	&	20.9	&	0	&	0.1	&	0	&	0.2	&	$< -15.0$	&	N	&	N	&	this work	\\
095330.53+562653.4	&	11452	&	2.0	&	3	&	0.01	&	0	&	0.01	&	$-14.0^{+0.4}_{-0.5}$	&	Y	&	N	&	D12	\\
105755.66+482502.0	&	11455	&	2.0	&	1	&	0.01	&	0	&	0.02	&	$< -13.8$	&	N	&	N	&	D12	\\
112637.74+513423.0	&	9234	&	4.7	&	1	&	0.05	&	0	&	0.02	&	$< -14.2$	&	N	&	N	&	Y14	\\
114343.76+550019.3	&	11460	&	1.8	&	0	&	0.0	&	0	&	0.01	&	$< -13.9$	&	N	&	N	&	D12	\\
115138.24+004946.4	&	7735	&	4.7	&	0	&	0.03	&	1	&	0.09	&	$< -14.2$	&	N	&	N	&	D09	\\
121629.13+601823.5	&	13860	&	24.0	&	21	&	0.09	&	5	&	0.2	&	$-14.1^{+0.1}_{-0.2}$	&	Y	&	Y	&	G14	\\
131926.52+105610.9	&	11470	&	1.8	&	5	&	0.03	&	2	&	0.0	&	$-13.6^{+0.3}_{-0.3}$	&	Y	&	Y	&	D12	\\
144052.60$-$023506.2	&	11474	&	1.8	&	0	&	0.03	&	1	&	0.07	&	$< -13.7$	&	N	&	N	&	D12	\\
144507.30+593649.9	&	7738	&	4.7	&	2	&	0.02	&	6	&	0.1	&	$-13.9^{+0.2}_{-0.3}$	&	Y	&	Y	&	D09	\\
153656.44+312248.1	&	11476	&	1.8	&	0	&	0.02	&	1	&	0.05	&	$< -13.7$	&	N	&	N	&	D12	\\
163159.59+243740.2	&	11483	&	1.9	&	24	&	0.05	&	8	&	0.02	&	$-12.9^{+0.1}_{-0.2}$	&	Y	&	Y	&	D12	\\
170246.09+602818.9	&	7739	&	4.7	&	0	&	0.03	&	2	&	0.1	&	$< -14.0$	&	N	&	N	&	D09	\\
233837.10$-$002810.3	&	5667	&	4.7	&	2	&	0.1	&	2	&	0.1	&	$-14.2^{+0.3}_{-0.5}$	&	Y	&	N	&	GH07c	
 \enddata
\tablecomments{
Column (1) galaxy name.
Column (2) \textit{Chandra} observation ID.
Column (3) effective exposure time of \textit{Chandra} observation.
Column (4) total number of soft (0.5-2~keV) X-ray counts within a $1\farcs$5 circular aperture centered on the optical source position (no background subtraction).
Column (5) expected number of soft (0.5-2~keV) background counts within each source aperture, estimated from background annuli with inner and outer radii of 5 and 10$\arcsec$, respectively.
Column (6) total number of hard (2-8~keV) X-ray counts.
Column (7) expected number of hard (2-8~keV) background counts.
Column (8) logarithm of the X-ray flux over the full 0.5-8~keV band,  assuming an absorbed power-law in PIMMS, with $\Gamma=2.0$ and $\nh$ set to the Galactic value along the line of sight.  A 92\% aperture correction (calculated at 2.3~keV) is applied to the net count rates for the flux calculation, to account for the fraction of source photons missed by our choice of $1\farcs5$ aperture.  If a source is not detected within the 0.5-8~keV band, then flux upper limits are reported at the 95\% confidence level.
Column (9) flag denoting whether an X-ray source is detected (Y) or not detected (N) at the 95\% confidence level over the full 0.5-8~keV band (according to \citealt{kraft91}).
Column (10) same as column (9) above, but at the 99\% confidence level.
Column (11) reference of paper that originally published each \textit{Chandra} observation.  \textit{D09}--\citet{desroches09}; \textit{D12}--\citet{dong12a}; \textit{G14}--\citet{gultekin14}; \textit{GH07c}--\citet{greene07}; \textit{Y14}--\citet{yuan14}.
}
\label{tab:xlimphot}
\end{deluxetable*}

\begin{deluxetable*}{l c c c c c c c c c c}
\tablecaption{X-ray Stacking Analysis}
\tablecolumns{11}
\renewcommand\arraystretch{1.2}

\tablehead{
		\multicolumn{3}{c}{}    &
		\multicolumn{3}{c}{Soft (0.5-2 keV)}  &
		\multicolumn{3}{c}{Hard (2-8 keV)}   &
		\multicolumn{2}{c}{} \vspace{0.2cm} \\ 
                 \colhead{Stack}       	  			&  
		\colhead{$N_{\rm src}$}			& 
		\colhead{$\tau_{\rm tot}$}   		&  
		\colhead{$C_{s,{\rm tot}}$}		& 
		\colhead{$B_s$}                		& 
		\colhead{$R_s$}                            & 
		\colhead{$C_{h,{\rm tot}}$}                 & 
		\colhead{$B_h$}                                 & 
		\colhead{$R_h$}                            & 
		\colhead{$\Gamma_{{\rm fixed}\, N_{\rm H}}$}                                   & 
		\colhead{${\log N_{\rm H}}_{{\rm fixed}\, \Gamma}$}                                    \\ 
                \colhead{}       	  		&  
		\colhead{}					 	& 
		\colhead{(ks)}					&  
		\colhead{(counts)}				& 
		\colhead{(counts)}  				& 
		\colhead{(counts s$^{-1}$)}	   	& 
		\colhead{(counts)}  				& 
		\colhead{(counts)}  				& 
		\colhead{(counts s$^{-1}$)}	& 
		\colhead{}						& 
		\colhead{(cm$^{-2}$)}				\\ 
		\colhead{(1)}					& 
		\colhead{(2)}					& 
		\colhead{(3)}					& 
		\colhead{(4)}					& 
		\colhead{(5)}					& 
		\colhead{(6)}					& 
		\colhead{(7)}					& 
		\colhead{(8)}					& 
		\colhead{(9)}					& 
		\colhead{(10)}					& 
		\colhead{(11)}					\\ 
}
\startdata
95\% conf.	&	12	&	77.9	&	5	&	0.5	&	$6.1\times10^{-5}$	&	6	&	1.0	&	$7.3\times10^{-5}$	&	$0.6^{+1.6}_{-0.6}$	&  22.1($<$22.3)\\
99\% conf.	&	14	&	84.6	&	10	&	0.6	&	$1.1\times10^{-4}$	&	8	&	1.1	&	$9.3\times10^{-5}$	&	1.0$^{+1.1}_{-0.5}$  &  21.8($<$22.1) \\
 \enddata
\tablecomments{
Column (1) subset of objects included in stacking analysis (non-detections at either the 95\% or 99\% confidence level).
Column (2) number of objects included.
Column (3) total stacked exposure time.
Column (4) total number of stacked soft (0.5-2~keV) X-ray counts within a $1\farcs$5 circular aperture centered on the optical source position (no background subtraction).
Column (5) expected number of stacked soft (0.5-2~keV) background counts within each source aperture.
Column (6) stacked net soft count rate (0.5-2~keV).  A 95\% aperture correction (calculated at 1~keV) is applied to account for the fraction of source photons missed by our choice of $1\farcs5$ aperture. 
Column (7) total number of stacked hard (2-8~keV) X-ray counts.
Column (8) expected number of  stacked hard (2-8~keV) background counts within each source aperture.
Column (9) stacked net hard count rate (2-8~keV).  An 88\% aperture correction (calculated at 4.5~keV) is applied to account for the fraction of source photons missed by our choice of $1\farcs5$ aperture.
Column (10) estimated $\Gamma$ that can explain the stacked hardness ratio from the stacked signals, assuming an absorbed power-law, and keeping the column density frozen at $\nh=2 \times 10^{20} {\rm cm}^{-2}$ (the average Galactic column density along each line of sight, weighted by the exposure time of each observation).  
Column (11) estimated $\log \nh$ that can explain the stacked hardness ratios, assuming an absorbed power-law with $\Gamma=2$.   The lower-limit on the 90\% confidence interval is unconstrained by the data, so in lieu of error bars, we   report the 90\% confidence upper limit in parentheses for this column.  
}
\label{tab:xstack}
\end{deluxetable*}

Next, we perform a stacking analysis on the 12 sources that remain undetected at the 95\% level, and on the 14 sources not detected at the 99\% level.  The stacked signals are presented in Table~\ref{tab:xstack}.  For both subsets, we obtain X-ray detections in both the soft and hard energy bands at $>$99\% confidence.   To confirm that the stacked detection is not an artifact  of improper background estimation \citep[see, e.g.,][]{willott11, cowie12} we perform the following test.  For the stack of 99\% non-detections (14 objects), we blindly displace the center of each image's source extraction region by $3\farcs$5, in a randomly chosen direction for each image (the magnitude of this offset is chosen to avoid overlap with each background extraction region and the original source  region).  We then repeat the stacking analysis. We find only one stacked count in the soft band, and one stacked count in the hard band, which is consistent with the expected background levels of 0.6 soft and 1.1 hard counts.

Finally, we use the observed hardness ratio of the stacked signals (defined by $R_h$/$R_s$, where $R_h$ and $R_s$ are the net count rates in the hard and soft bands, after incorporating 88 and 95\% aperture corrections, respectively, appropriate at 1 and 4.5~keV for $1\farcs5$ extraction regions) to explore the average spectral properties of the X-ray non-detected objects.  We first assume an absorbed power-law in PIMMS, with the  column density fixed to $\nh=2\times10^{20}~{\rm cm}^{-2}$, which is the  (exposure-time weighted) average of the Galactic $\nh$ values for each source; the range of $\Gamma$ that can explain the observed hardness ratios of the stacked samples are reported in Column (10) of Table~\ref{tab:xstack}.  We also perform a similar test fixing $\Gamma=2$ to identify the range of $\nh$ that can replicate the observed hardness ratios (see column 11 of Table~\ref{tab:xstack}).  The uncertainties on $\Gamma$ and $\nh$ estimated in this manner are large, and we cannot break the degeneracy between column density and photon index with so few stacked counts.  Still, we can confidently assert from this analysis that the population of X-ray non-detected \imbh\ AGN have, on average, a relatively hard observed X-ray spectrum (either caused by a photon index flatter than most of the X-ray detected objects, by a modest amount of intrinsic absorption, or a combination of the two).

\section{Results}
\label{sec:res}

\subsection{Nuclear X-ray Emission from  SMBHs}
\label{sec:nuclear}

\begin{figure}
\includegraphics[scale=0.42]{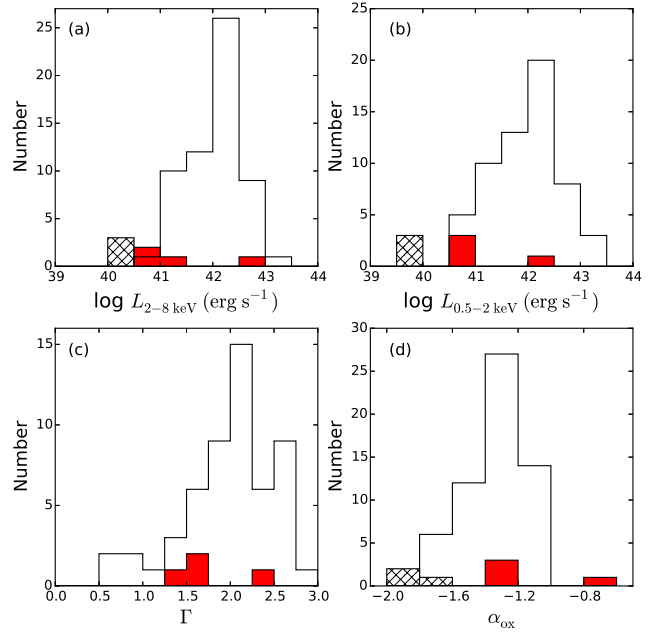}
\caption{X-ray properties of our seven lower-Eddington ratio \textit{Chandra} targets (red histograms), with cross-hatched histograms indicating upper limits.  The X-ray properties of the \parent\ sample are  shown for reference (open histograms).  For clarity, only \parent\ objects with X-ray detections are included.   (a) Hard X-ray luminosity from 2-8~keV.  (b) Soft X-ray luminosity from 0.5-2~keV.  (c) Best-fit photon spectral index $\Gamma$ (only four \textit{Chandra} targets with X-ray detections are shown).  (d) X-ray to UV luminosity ratio parameterized by $\alpha_{\rm ox}$.  More negative numbers are ``X-ray weaker.''}
\label{fig:xray}
\end{figure}

We associate all four X-ray detections with nuclear \imbh s, as described below.  All four sources have hard X-ray luminosities $\lx \gtrsim 10^{41}~\ergs$.  These X-ray luminosities are as expected given the Eddington ratios derived by \citetalias{greene07b}  for our \textit{Chandra} targets: for $\lledd > 10^{-2}$ and $\mbh = 10^6~\msun$ (see Table~\ref{tab:obslog}), we expect $\lx > 10^{41}~\ergs$ if 10\% of the bolometric luminosity is emitted in the hard X-ray band.    Excluding \srcseven, both the soft and hard X-ray luminosities of our  \textit{Chandra} targets generally populate the faint end of  the \parent\ \citetalias{greene07b} \imbh\ sample (Figure~\ref{fig:xray}).  For completeness, we  show the best-fit photon indices for the four X-ray detections in Figure~\ref{fig:xray}c, which cover a similar range as the \parent\ sample.  

It is very unlikely that a stellar mass black hole (or even multiple stellar mass black holes within the \textit{Chandra} point spread function) could produce the observed amount of X-ray emission from each galaxy.  The high-luminosity tail of a galaxy's X-ray binary (XRB) population contains the vast majority of known ultraluminous X-ray sources (ULXs; i.e., most ULXs are stellar mass black holes accreting at super-Eddington rates; e.g., \citealt{begelman02}).  Accounting for a combination of super-Eddington accretion and mild beaming, it is difficult for a stellar mass black hole to attain an X-ray luminosity $>$10$^{41}~\ergs$ \citep{poutanen07}.  The  largest ULX catalogs  observationally confirm that objects with $\lx >10^{41}~\ergs$ are extremely rare and very unlikely attributed to luminous XRBs \citep[][also see \citealt{feng11} for a  review]{swartz11, walton11}.  Furthermore, the brightest ULXs are typically found in galaxies with high specific star formation rates, which contain  (short-lived) high-mass X-ray binary populations \citep[e.g.,][]{king02, grimm03, mineo14}.  ULXs in galaxies with little star formation arise from low-mass XRBs (likely applicable to 3/4 of our galaxies with X-ray detections, as judged from their SDSS optical colors; see Section~\ref{sec:optical}), and they appear to always have $\lx < 2\times10^{39}~\ergs$ \citep{irwin04}.  ULXs are also $\sim$10 times less common in such  galaxies \citep[e.g.,][]{swartz11, walton11, plotkin14}, since the number of low-mass XRBs scales with the total stellar mass instead of the specific star formation rate \citep{gilfanov04}.  Thus, the presence of hard X-rays at $>$10$^{41}~\ergs$ strongly advocates that our four X-ray detected galaxies  host  AGN.

\subsection{X-ray to UV Luminosity Ratios}
\label{sec:aox}

One of the most common ways to quantify the relative amount of accretion power outputted in the X-ray waveband is through the $\alpha_{\rm ox}$ parameter, which measures the ratio  of  X-ray to UV luminosity  \citep{tananbaum79}.   We adopt $\alpha_{\rm ox} = -0.3838 \log \left(\luv/\lkev\right)$,  where $\lkev$ and $\luv$ are (unabsorbed) X-ray and UV luminosity densities at rest-frame 2~keV and 2500~\AA, respectively.  We use PIMMS and the adopted spectral parameters for each source (see Section~\ref{sec:xrayspec}) to calculate $\lkev$.  For consistency with \citetalias{dong12}, $\luv$ calculations are based on broad H$\alpha$ line luminosities ($L_{\rm H\alpha}$) as follows.  First, we take the \citetalias{greene07b} $L_{\rm H\alpha}$ measurements and use the relation  $L_{\rm H\alpha} = 5.25\times10^{42}\left({\rm L_{5100}}/10^{44}\ \ergs\right)^{1.157} \ergs$ \citep{greene05} to determine the  luminosity  of the AGN continuum at 5100~\AA\ ($L_{\rm 5100}$), and we then assume that the continuum follows a power-law of the form $f_{\nu} \propto \nu^{-0.44}$ \citep{vanden-berk01} to estimate  $\luv$.    We list  $\alpha_{\rm ox}$ values in Table~\ref{tab:xrayspec}, and the $\alpha_{\rm ox}$ distribution is shown in Figure~\ref{fig:xray}d.   

\citet{desroches09} and \citetalias{dong12} show that the \parent\ sample has on average harder (i.e., less negative) $\alpha_{\rm ox}$ values compared to quasars with larger black hole masses (i.e., \imbh s are systematically ``X-ray brighter'').   \citetalias{dong12} explain this as being driven primarily by black hole mass, caused by lower-mass black holes being fed by higher temperature accretion disks.  \citetalias{dong12} find evidence for a weak  anti-correlation between $\alpha_{\rm ox}$ and $\mbh$ among the \parent\ sample, which supports their interpretation (an anti-correlation between $\alpha_{\rm ox}$ and $\log \mbh$ has similarly been observed among quasars; e.g., \citealt{kelly08}). However, \citetalias{dong12} also show that $\alpha_{\rm ox}$ is not as hard as expected, if one were to extrapolate from the well-known anti-correlation between $\alpha_{\rm ox}$ and  $\luv$ that is defined by more luminous (and massive) quasars \citep[e.g.,][]{avni82, steffen06, just07}.    A potential flattening of the $\alpha_{\rm ox} - \luv$ relation at low luminosities was also hinted at by \citet{steffen06}, and  seen by \citet{maoz07} as well for a sample of 13  low ionization nuclear emission line region (LINER) galaxies.

While most of our \textit{Chandra} targets are  also X-ray brighter than luminous quasars, they tend to be X-ray weaker (on average) than expected for their UV luminosities --- excluding \srcseven, the six other targets have $\Delta \alpha_{\rm ox}  = \alpha_{\rm ox} - \alpha_{\rm ox, exp} < -0.3$ (see Table~\ref{tab:xrayspec}), where $\alpha_{\rm ox, exp}$ is the average $\alpha_{\rm ox}$ expected from the \citet{just07} $\alpha_{\rm ox} - \luv$ relation.  Our targets' $\Delta \alpha_{\rm ox}$  values correspond to being X-ray weaker than $\alpha_{\rm ox, exp}$ at the $3.6-8.5$$\sigma$  level (see Table~\ref{tab:xrayspec}; the rms  $\sigma$ deviation for each object is tabulated from \citealt{steffen06}).   Finally, we note that \srcseven\ is the X-ray brightest \imbh\ from \citetalias{greene07b} observed so far, but it does not have an unusual X-ray spectrum or column density that would indicate that it is fundamentally different than the other \citetalias{greene07b} objects.

  \subsection{Narrow Line Emission and AGN Classification}
 \label{sec:optical}
 All \ntarg\ \textit{Chandra} targets are formally below the broad line detection thresholds of \citetalias{greene07b}, who primarily require  $f_{{\rm H}\alpha}/\sigma_{\rm rms}>200$ and EW$\left[{\rm H}\alpha\right]>15$~\AA\ ($f_{{\rm H}\alpha}$ is the H$\alpha$ line flux, $\sigma_{\rm rms}$ is the rms deviation of the continuum subtracted spectrum, and EW$\left[{\rm H}\alpha\right]$ is the H$\alpha$ equivalent width; see \citealt{greene07a} for details).  Our targets were included in the  \citetalias{greene07b} sample because visual inspection of their spectra (by \citetalias{greene07b}) revealed broad and interesting H$\alpha$, and  they were subsequently flagged by \citetalias{greene07b} as less secure ``candidate'' black holes (which they term as their \textit{c} sample). We  assess our targets' AGN identifications here by re-examining their optical properties  in conjunction with our new \textit{Chandra} X-ray constraints.
 
 As an initial cross-check, we consult the results of an independent optical analysis by \citet{reines15}, who analyzed the SDSS spectra of $\sim$67000 emission line galaxies.  \citet{reines15} detected broad H$\alpha$ in all of our targets except for \srcone.\footnote{Note that not all of our targets are included in the final 262 object AGN sample of \citet{reines15}, as they require both broad H$\alpha$ and Seyfert-like narrow emission lines.}  
  We therefore operate under the assumption that only for \srcone\ could the broad H$\alpha$ seen by \citetalias{greene07b} during visual inspection be a statistical ``false positive.''  That is, we consider the claim for the presence of broad H$\alpha$ in these objects to be robust.  We stress, however, that broad H$\alpha$ on its own does not prove that an AGN is present, especially in star forming galaxies where there could be sources of contamination from young stars and/or supernovae \citep[see e.g.,][Baldassare et al.\ subm.]{filippenko89, greene04, reines13}.

\begin{figure}
\includegraphics[scale=0.68]{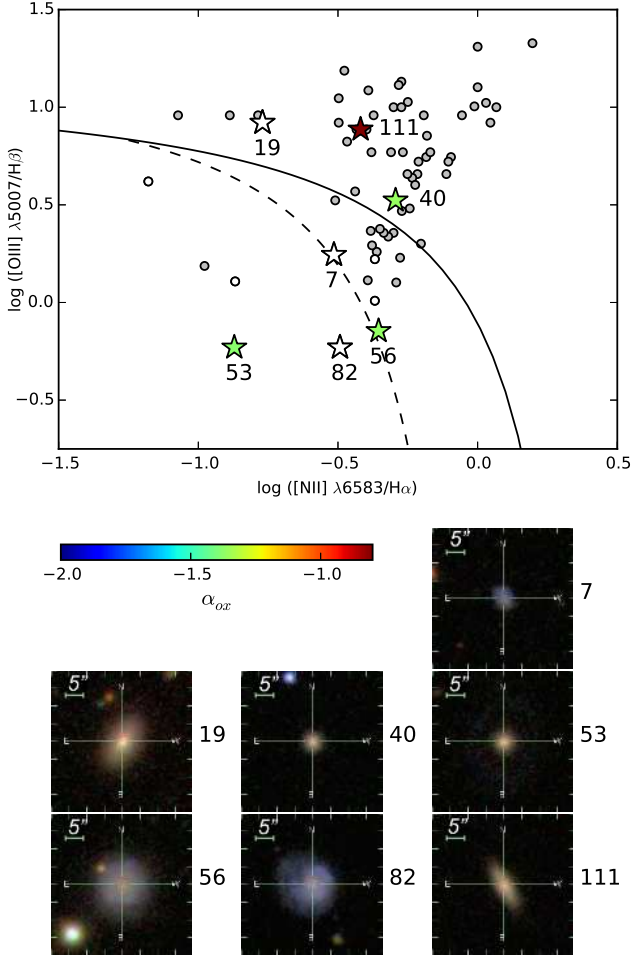}
\caption{Top panel: BPT diagram showing narrow emission line ratios  for our seven \textit{Chandra} X-ray targets (star symbols), color coded by X-ray brightness $\alpha_{\rm ox}$  (see color bar).  Open star symbols denote X-ray non-detections.  The solid curve shows the ``maximum starburst line'' from \citet{kewley06}, derived from pure stellar photoionization models.  Galaxies above the solid curve are ``Seyfert-like.''  The dashed curve shows the empirical dividing line between star forming and active galaxies \citep{kauffmann03}.     AGN  from the \parent\ sample are overplotted for comparison (circles), with open symbols denoting X-ray non-detections.  The bottom panels show SDSS $gri$ color composite images of each of our \textit{Chandra} targets.  The green scale bar at the top left of each image represents 5$\arcsec$. }
\label{fig:bpt}
\end{figure}

In Figure~\ref{fig:bpt} we show the location of each of our targets in the Baldwin-Phillips-Terlevich (BPT) diagram \citep[e.g.,][]{baldwin81, kewley01, kewley06, kauffmann03}, which uses narrow emission line diagnostics to separate purely star forming galaxies from galaxies with harder (AGN-like) ionizing continua.  For reference, we  include the \parent\ sample in Figure~\ref{fig:bpt}.   Three of our \textit{Chandra} targets have narrow optical line ratios typical of Seyfert galaxies (galaxies \srcthreedig, \srctwodig, and \srcsevendig).   These three galaxies also appear to have red SDSS optical colors (see bottom panels of Figure~\ref{fig:bpt}), which suggests negligible star formation (although, see \citealt{jiang11} for a high spatial resolution morphological study  that indicates that all three galaxies contain a disk component).  The combination of the narrow line diagnostics and the presence of broad H$\alpha$ emission highly suggests that all three of these galaxies host an AGN.    Galaxies \srctwodig\ and \srcsevendig\ also display  X-ray emission that confirms their AGN nature (see Section~\ref{sec:nuclear}).  We note that the absence of X-ray emission is not sufficient to exclude the presence of an AGN.    Although we do not detect X-rays from galaxy \srcthreedig, we  have a sensitive limit ($\alpha_{\rm ox}<-1.72$), which places it in the X-ray weak tail defined by the \parent\  sample \citepalias[][also see Section~\ref{sec:disc}]{dong12}.

 None of the other four galaxies show obvious optical narrow-line AGN signatures in Figure~\ref{fig:bpt}.   However, two of these galaxies are  still very likely AGN based on their X-ray properties (galaxies \srcfourdig\ and \srcfivedig; we note that \citealt{dong12} independently classify \srcfour\ as an \imbh\ AGN from broad H$\alpha$ in its optical SDSS spectrum).  Galaxies \srconedig\ and \srcsixdig\ do not show X-ray emission or AGN-like narrow line ratios,  they have blue SDSS optical colors, and it is unclear if \srcone\ displays broad H$\alpha$.   We are therefore uncertain on the AGN nature for galaxies \srconedig\ and \srcsixdig, and we consider them to be low-confidence AGN \textit{candidates} here.  We note that  \srcone\ is not included in the low-$\mbh$ sample of \citet{dong12}, but \srcsix\ is included in their sample.  
 
Among the six  \imbh s with low-$\lledd$ for which we take X-ray information from the literature, the two from \citet{gultekin14} were also flagged by \citetalias{greene07b} as part of their \textit{c} sample.  Following similar arguments as above, we consider both sources to be AGN here: both have AGN-like $\log$~[\ion{O}{3}]/H$\beta > 1.3$ measured by \citetalias{greene07b}, and nuclear X-ray point sources.   \citet{yuan14} discuss classification of their four targets in detail, and we consider all four of their objects to be  bona fide \imbh\ AGN.

  In summary, we consider galaxies \srctwodig\ and \srcsevendig\ to be AGN based on their optical and X-ray properties, and \srcthree\ to be an X-ray weak AGN.  Galaxies \srcfourdig\ and \srcfivedig\ appear to be AGN that lack AGN-like narrow-emission lines, and we consider galaxies \srconedig\ and \srcsixdig\ to be (lower-confidence) AGN candidates.  

 \subsection{Bolometric Luminosities and Eddington Ratios}
 \label{sec:lbol}
Throughout the text, we generally adopt the Eddington ratios calculated by \citetalias{greene07b} for our \textit{Chandra} targets.  For these calculations, a bolometric correction of $\lbol = 9.8 L_{\rm 5100}$ \citep{mclure04} is adopted, where $L_{5100}$ is the luminosity of the AGN continuum at 5100~\AA\ (which is estimated from the broad H$\alpha$ line luminosity; see Section \ref{sec:aox}).  We adopt the \citetalias{greene07b} $\lledd$ estimates here to ease comparison with the literature, as  $\lledd$ estimates for the \parent\ sample are also drawn from \citetalias{greene07b}.    It is possible that  this specific choice of bolometric correction could systematically over- or under-estimate $\lledd$.  However, such a bias would affect the the entire sample in the same direction, on average.  Therefore, by calculating $\lledd$ in a consistent manner between all subsets, we can reliably search for trends as a function of (relative) Eddington ratio, as long as we bear in mind the possibility for a systematic offset for the entire sample when interpreting the results.

At the moment, bolometric corrections for \imbh\ AGN  are still poorly constrained, and  empirical constraints  will require multiwavelength, high-spatial resolution imaging of  the nuclear emission for a sizable sample of objects.  In the meantime, we explore three other methods of  applying bolometric corrections to estimate $\lbol$ for AGN, which are described below and summarized in Table~\ref{tab:lledd}:

 \begin{enumerate}
 
 \item $k(\mbh)=\lbol/L_{5100}$:  \citetalias{dong12} suggest that the bolometric correction could include a  \bh\ mass term, such that  $\log \left(L_{\rm bol}/L_{5100}\right) = -0.54 \log \mbh + 5.43$ (see their Equation 3).  For a 10$^6~\msun$ \imbh, this leads to $L_{\rm bol}=155 L_{5100}$,  which  increases the  estimates from \citetalias{greene07b} by $\approx$1.2 dex (see column 3 of Table~\ref{tab:lledd}).

 \item $k_1 = \lbol/L_{5100}$: we also estimate $\lledd$ from the strength of [\ion{O}{3}]  line  emission in the SDSS optical spectra.  Following \citet{kauffmann09}, we assume $\lbol \approx 500-800$ $L$[\ion{O}{3}].  When calculating $L$[\ion{0}{3}], we include  a correction for extinction by estimating $E(B-V)$ from the  Balmer-decrement in the SDSS spectra (assuming a \citealt{cardelli89} extinction curve and $L_{\rm H\alpha}$/$L_{\rm H\beta}$ = 3.1, which is typically adopted for AGN to include photoionization + collisional enhancement of Hydrogen lines; e.g., \citealt{halpern83}). Since this $\lbol$ estimate assumes photoionization by an AGN continuum, we only perform the calculation for the three targets with AGN-like narrow emission line ratios in Figure~\ref{fig:bpt} (galaxies \srcthreedig, \srctwodig, \srcsevendig; see column 4 of Table~\ref{tab:lledd}).

 \item $k_2 = \lbol / L_{\rm 2-10 keV}$: we also estimate $\lledd$ from the observed X-ray luminosities.  Following \citet[][see their Section~5]{yuan14}, we apply an X-ray bolometric correction for \imbh s of $\lbol = 7-20\ L_{\rm 2-10 keV}$ (see column 5 of Table~\ref{tab:lledd}). 
 
 \end{enumerate}
  
Both the mass-dependent  and $L$[\ion{O}{3}]-based bolometric corrections  suggest more rapidly accreting AGN than estimated by \citetalias{greene07b}.  The X-ray based bolometric corrections generally predict lower Eddington ratios (except for \srcseven, the lone object with a positive $\Delta \alpha_{ox}$ value).   The discrepancy between  different methods of calculating $\lledd$ serves as a quantitative guide for the degree to which systematics may affect our adopted $\log \lledd$ estimates.   We stress, however, that the discrepancies between different $\lbol$ estimates do not imply that the \citetalias{greene07b} values must be incorrect or biased.  Rather, the discrepancies  motivate the need for high-resolution broadband imaging of the nuclei of a large sample of \imbh\ AGN, to provide observational constraints to calibrate bolometric corrections against.

\begin{deluxetable}{l c c c c}
\tablecaption{$\log \lledd$ Using Different Bolometric Corrections}
\tablecolumns{5}
\renewcommand\arraystretch{1.2}
\tablehead{
	       \colhead{}    &
	      \multicolumn{4}{c}{$\log \lledd$} \vspace{0.1cm}  \\
                \colhead{GH07 ID}       	  		&  
		\colhead{9.8$L_{\rm 5100}$}		& 
		\colhead{$k\left(M_{\rm BH}\right) L_{\rm 5100}$}		&  
		\colhead{$k_1$L[\ion{O}{3}]}		& 
		\colhead{$k_2 L_{\rm 2-10keV}$}  \\ 
		\colhead{(1)}					& 
		\colhead{(2)}					& 
		\colhead{(3)}					& 
		\colhead{(4)}					& 
		\colhead{(5)}					
}
\startdata
7        &     $-$1.7 &     $-$0.4 &  \nodata        & $<$$-$2.2         \\
19       &     $-$1.6 &     $-$0.5 & $-$0.8 -- $-$0.6     & $<$$-$2.6         \\
40       &     $-$1.5 &     $-$0.3 & $-$1.1 -- $-$0.8     & $-$2.1 -- $-$1.6     \\
53       &     $-$1.8 &     $-$0.7 &  \nodata        & $-$2.6 -- $-$2.2     \\
56       &     $-$2.0 &     $-$0.9 &  \nodata        & $-$2.5 -- $-$2.1     \\
82       &     $-$1.6 &     $-$0.4 &  \nodata        & $<$-2.4         \\
111      &     $-$1.5 &     $-$0.4 & $-$0.5 -- $-$0.3     & $-$0.5 -- $-$0.0     \\

 \enddata
\tablecomments{
Column (1) \citetalias{greene07b} galaxy ID.
Column (2) $\log \lledd$ from \citetalias{greene07b} (repeated from Table~\ref{tab:obslog}), assuming $\lbol = 9.8 L_{5100}$.
Column (3) $\log \lledd$ assuming $\lbol = 10^{5.43} \mbh^{-0.54} L_{\rm 5100}$ (see \citetalias{dong12}).
Column (4) range of $\log \lledd$, assuming $\lbol=k_1L$[\ion{O}{3}], where $k_1$=500-800 (see \citealt{kauffmann09}).  $L$[\ion{O}{3}] is calculated by applying a Balmer-decrement based correction for extinction to the [\ion{O}{3}] line fluxes presented by \citetalias{greene07b}. Values are presented only for our three \textit{Chandra} targets displaying narrow-line emission  indicative of photoionization by an AGN.
Column (5) range (or limits) of $\log \lledd$ assuming $\lbol=k_2 L_{\rm 2-10 keV}$, where $k_2=7-20$, and $L_{\rm 2-10 keV}$ is the X-ray luminosity (or limit) determined by our \textit{Chandra} observations.}
\label{tab:lledd}
\end{deluxetable}

\section{Discussion}
\label{sec:disc}
 
\begin{figure}
\includegraphics[scale=0.5]{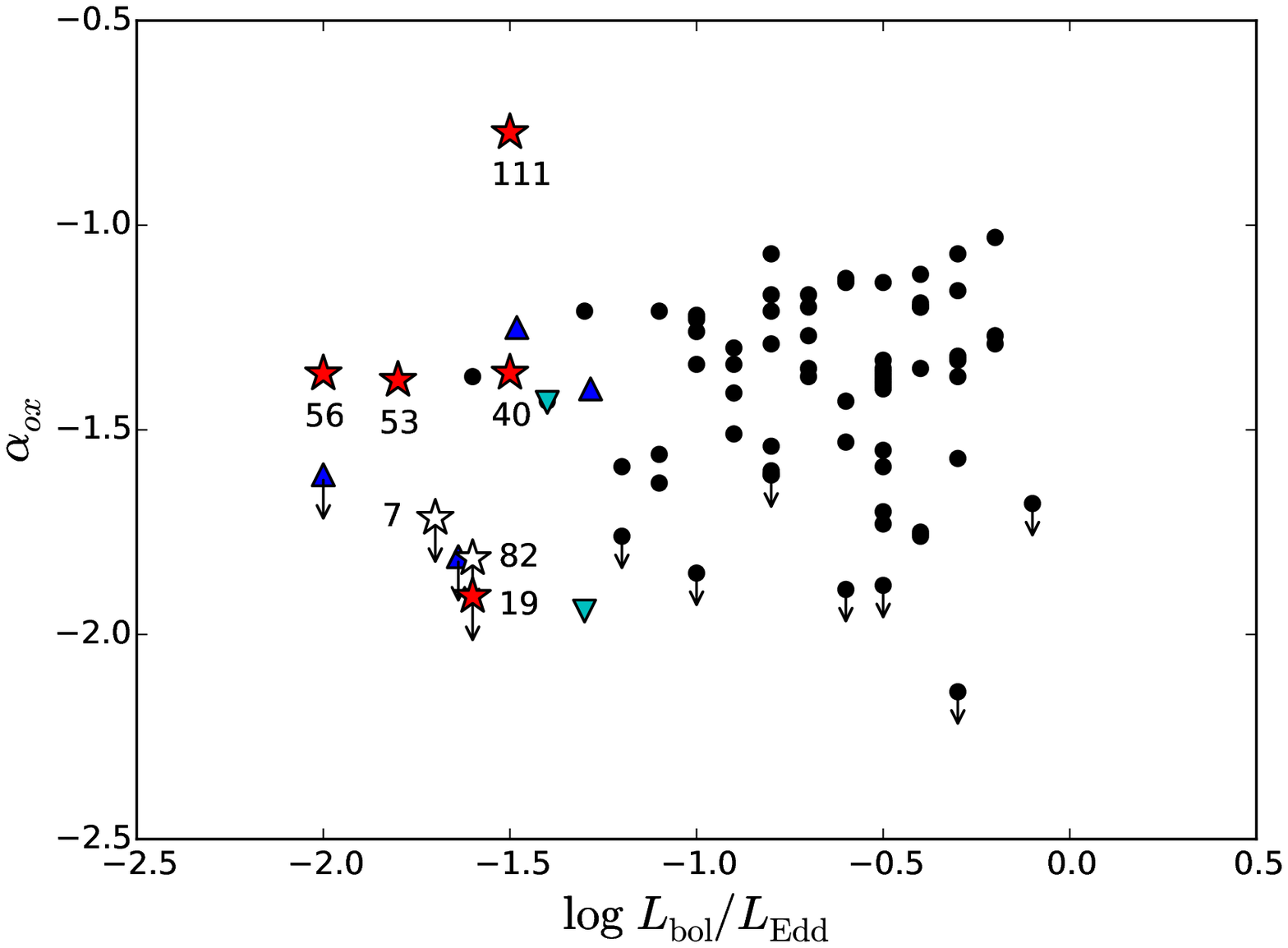}
\caption{$\alpha_{\rm ox}$ as a function of Eddington ratio for our  \textit{Chandra} targets (red filled star symbols), the four \imbh\ AGN with $\log \lledd < -1$ from \citet[][blue triangles]{yuan14}, the two \imbh\ AGN with $\log \lledd < -1$ from \citet[][cyan upside down triangles]{gultekin14}, and  the \parent\ sample (circles).  Arrows denote upper limits on $\alpha_{\rm ox}$.  The open star symbols represent our two less confident AGN candidates (galaxies \srconedig\ and \srcsixdig; see Section~\ref{sec:optical}).}
\label{fig:aox}
\end{figure}

 Here, we compare the X-ray properties of  the lower-Eddington ratio AGN to the \parent\ sample (differences between the X-ray properties of accreting \imbh s vs.\ \smbh s have already been explored in depth by \citetalias{dong12}; also see \citealt{greene07, desroches09}).     We generally exclude the two lower-confidence AGN candidates (galaxies \srconedig\ and \srcsixdig) from the following discussion, unless stated otherwise.  Our low-$\lledd$ sample therefore contains 11 objects, and the \parent\ comparison samples contains 67 objects.  As noted in Section~\ref{sec:lbol}, we adopt the $\lledd$ estimates from \citetalias{greene07b}, as that provides the most straightfoward method for  uniformly comparing $\lledd$ across the entire 78 object sample.  While these estimates may be systematically offset from the ``true'' $\lledd$ values, the estimates provide a  reliable tracer for the \textit{relative} Eddington ratios between objects in the full sample.  We cannot calculate $\lledd$ from $L$[\ion{O}{3}] across the entire sample because not all objects show narrow-line AGN signatures.  The full 78 object sample  shows a large dispersion in $\alpha_{\rm ox}$ (see Section \ref{sec:wlq}), which implies that applying an X-ray based bolometric correction for determining $\lledd$ is not straightforward, and would likely require an undetermined correction that is dependent  on $\alpha_{\rm ox}$.   
  
  \citetalias{dong12} report on the lack of a correlation between $\alpha_{\rm ox}$ and $\log \lledd$ within just the \parent\ sample (also see \citealt{greene07, desroches09}).   After extending the dynamic range in $\log \lledd$ by almost an order of a magnitude, we  do not see any  correlation either (Figure~\ref{fig:aox}).   We  perform a linear regression on $\alpha_{\rm ox}$ vs.\ $\log \lledd$ (including upper limits on $\alpha_{\rm ox}$; \citealt{kelly07}), and we find a slope consistent with zero ($0.1\pm0.1$).  Furthermore,  generalized Kendall's $\tau$ and Spearman's $\rho$ correlation tests  on the combined 78 object sample both indicate that no correlation is present at the $p=0.1$ level.  We incorporate upper limits on the X-ray non-detections when running these correlation tests by using the Astronomy SURVival Analysis (ASURV) package rev 1.2 \citep{lavalley92}, which  implements  the statistical methods presented in \citet{feigelson85}.  The lack of a correlation within just the 67 object \parent\ sample reported by \citetalias{dong12} is  therefore not solely due to their limited dynamic range.

In addition, we do not see any evidence for statistically different X-ray properties between the low- and \parent\ samples.  A Peto-Prentice test (run through ASURV to incorporate upper limits)  indicates that the low- and high-Eddington ratio  samples do not follow statistically different distributions in $\alpha_{\rm ox}$  ($p=0.2$; also see Figures \ref{fig:xray}d and \ref{fig:aox}).    We also estimate the average $\alpha_{\rm ox}$ values for each distribution  using the Kaplan-Meier estimator in ASURV.  The 11 lower-$\lledd$ AGN have a mean $\left<\alpha_{\rm ox}\right>=-1.5\pm 0.1$, which is comparable to   $\left<\alpha_{\rm ox}\right>=-1.4\pm 0.04$ for the 67 object high-$\lledd$ sample.  Furthermore, for our four \textit{Chandra} objects to which we could fit a spectrum, we similarly do not see any meaningful differences in the spectral properties between the two samples (Figure~\ref{fig:xray}c).   We might have expected to see a correlation between (hard X-ray) $\Gamma$ and $\log \lledd$, as is observed for luminous quasars with $\log \lledd \gtrsim -2$ \citep[e.g.,][]{shemmer08, risaliti09, brightman13, ricci13}, but  proper investigation will require a sample with tighter X-ray spectral constraints, and thus deeper X-ray observations.

\begin{figure}
\includegraphics[scale=0.45]{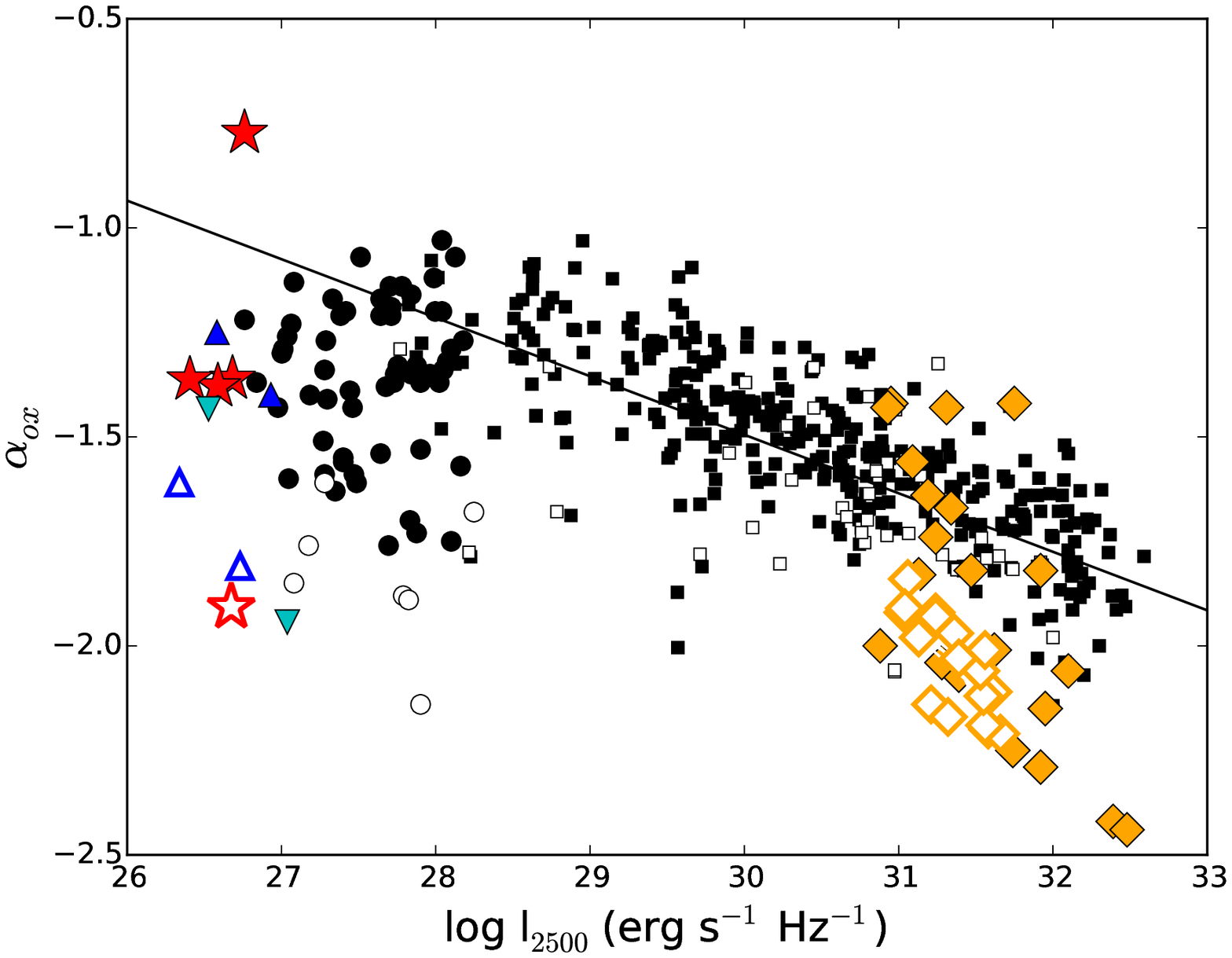}
\caption{$\alpha_{\rm ox}$ as a function of $\luv$ for our  low-$\lledd$ \textit{Chandra} targets (red star symbols; excluding the less-confident AGN in galaxies \srconedig\ and \srcsixdig), the four \citet{yuan14} low-$\lledd$ objects (blue triangles),  the two \citet{gultekin14} low-$\lledd$ objects (cyan upside down triangles), and the 67 objects from the \parent\ \imbh\ sample (circles).   For comparison, we plot samples of more massive black holes, including WLQs (at $z>1.5$; orange diamonds),  ``normal'' type 1 SDSS quasars from \citet[][squares]{just07}, and the best-fit $\alpha_{\rm ox} - \log \luv$ relation from \citet[][solid line]{just07}.  All open symbols denote upper limits on $\alpha_{\rm ox}$.}
\label{fig:aoxwlq}
\end{figure}

\subsection{The X-ray weak tail}
\label{sec:wlq}
The lower-$\lledd$ \imbh s appear to show as wide of a spread in $\alpha_{\rm ox}$ as the \parent\ sample (see Figure~\ref{fig:aox}).   Given that very few \imbh s display  signs of X-ray absorption (and those that do typically have moderate column densities, $\nh \approx 10^{22}$~cm$^{-2}$), some \imbh s might be intrinsically X-ray weak.  They could be similar to the nearby (z=0.192) narrow-line Seyfert 1 galaxy PHL 1811, which may have a smaller or quenched X-ray corona \citep[e.g.,][]{leighly07a, leighly07}.   Intriguingly, the range in $\alpha_{\rm ox}$ displayed by  \imbh s is also reminiscent of the X-ray properties of weak emission line quasars (WLQs; see, e.g., \citealt{shemmer09, wu11, wu12, luo15}).  WLQs are higher-redshift (mostly known  at $z\gtrsim1$), unobscured, radio-quiet quasars that have unusually weak  high-ionization broad emission lines (especially Ly$\alpha$ and \ion{C}{4} $\lambda$1549; see, e.g., \citealt{fan99, diamond-stanic09, plotkin10a, plotkin10, shemmer10, lane11}).    Approximately 50\% of known WLQs are significantly X-ray weaker than expected for their UV luminosities (with  $\Delta \alpha_{\rm ox}<-0.2$; see \citealt{shemmer09, wu12, luo15}).

In Figure \ref{fig:aoxwlq}, we compare the $\alpha_{\rm ox}$ distribution of \imbh s to  WLQs  as a function of $\luv$, with typical SDSS Type 1 quasars from \citet{just07} plotted for reference.  We  include 38 WLQs  that were optically-selected  from the SDSS, that have  $z>1.5$ (to ensure SDSS spectroscopic coverage of the high-ionization emission line \ion{C}{4}), and that have  \textit{Chandra} X-ray observations presented by  \citet{wu11, wu12} and \citet{luo15}.\footnote{About 35\% of the WLQs shown in Figure~\ref{fig:aoxwlq} were initially selected as high-redshift analogs to PHL 1811, which are nearly always X-ray weak and appear to share many similarities to WLQs in their optical and UV spectra.  For convenience, we refer to all objects studied by \citet{wu11,wu12} and \citet{luo15} as WLQs.}  The lack of WLQs at $\luv<10^{31}~\ergs$~Hz$^{-1}$ in Figure~\ref{fig:aoxwlq} is largely (but unlikely solely) due to the restriction in redshift. 
  
 It is clear from Figure~\ref{fig:aoxwlq} that both the \imbh\ and WLQ populations display a larger dispersion in $\alpha_{\rm ox}$ compared to `normal' Type 1 SDSS quasars.    To quantify the  dispersion, we use  the Kaplan-Meier estimator.  Since we do not see any statistical difference in the X-ray properties between the low- and \parent\ \imbh\ samples,   we consider the entire 78 object \imbh\ sample in the following, in order to improve statistics.  We find that the $25^{th}$-75$^{th}$ percentiles of the $\alpha_{\rm ox}$ distributions for the 78 \imbh s and 38 WLQs span a range of $0.36\pm0.07$ and $0.57\pm0.10$, respectively (errors are the standard deviations on the $25^{th}$ and 75$^{th}$ percentile $\alpha_{\rm ox}$ values added in quadrature);   when limiting the comparison  \citet{just07} SDSS quasars to similar luminosities as the WLQ sample (120 Type 1 quasars with $\luv > 10^{31}~\ergs$~Hz$^{-1}$), luminous quasars have 25$^{th}$-75$^{th}$ percentiles spanning only $0.22\pm0.06$ in $\alpha_{\rm ox}$.   The larger dispersion hints at a potential difference in accretion properties between \imbh\ AGN and `normal' Type 1 quasars.    In the next subsection, we consider  an analogy with WLQs to explore one potential mechanism for the larger dispersion in $\alpha_{\rm ox}$, namely accretion via a slim disk. 
 
\subsection{Comparison to \citet{luo15} and Slim Disk Accretion}
\label{sec:luo15}

For $\lledd \gtrsim  0.1-0.3$, accretion is expected to take place in the ``slim disk'' regime \citep[e.g.,][]{abramowicz88, szuszkiewicz96, bonning07, straub11}, where advective cooling losses are comparable to radiative losses, and the accretion flow becomes geometrically thick and radiatively inefficient  \citep[see, e.g., Section 6 of][for a brief review]{abramowicz13}.  \citet{luo15} propose that the unusual X-ray properties of WLQs (i.e., the large fraction of X-ray weak objects) may be related to accretion via slim disks,  as described below.

\citet{wu11,wu12}  initially suggested that WLQs contain a column of X-ray shielding gas that is very local to the black hole (interior to the broad emission line region); this X-ray shielding gas can produce highly X-ray weak objects at certain orientations, and X-ray normal objects at other orientations. \citet{luo15} went on to physically associate this shielding gas with the inner edge of a (geometrically thick) slim disk.  In this picture,     all WLQs are  fed by an inner slim  disk, but only the ones oriented such that we are  viewing the central engine through the ``puffed up'' disk material appear to be X-ray weak.\footnote{We stress that shielding from a geometrically thick slim disk is just one potential explanation for the WLQ phenomenon.  Other ideas  range from quenched X-ray coronae, to evolutionary effects, to gas deficient and/or  multi-zone broad emission line regions (see, e.g., \citealt{leighly07, hryniewicz10, laor11, liu11, banados14, shemmer15, plotkin15}).} For these X-ray weak objects, the direct X-ray continuum will be highly absorbed, and any detected X-ray emission should be dominated by reflected/scattered light \citep{luo15}.  This scenario results in the population of WLQs as a whole displaying a large observed dispersion in $\alpha_{\rm ox}$.

It is tempting to appeal to a similar scenario to explain the large dispersion in $\alpha_{\rm ox}$ observed for  \imbh s.  However, if such a scenario were to apply to \imbh\ AGN, then we  expect to see a systematic change in the X-ray properties of the 78-object \imbh\ sample around $\lledd \approx 0.1-0.3$, as the accretion flow transitions between  a geometrically thick slim disk and a geometrically thin (radiatively efficient) \citet{shakura73} disk.  In particular, we should see less dispersion in $\alpha_{\rm ox}$ at $\lledd \lesssim 0.1-0.3$, in the context of a scenario where X-rays at these lower Eddington ratios originate from inverse Compton scattering of disk UV photons off a hot corona that is energetically coupled to a thin  disk \citep[e.g.,][]{haardt93}.  For $\lledd \lesssim 0.1-0.3$, the higher UV flux from the thin disk would increase the number of inverse Compton scatterings, thereby cooling the corona and also producing mildly steeper X-ray spectra on average \citep[e.g.,][]{ghisellini94}.

  A systematic change in the X-ray properties, as described above, is not observed among the \imbh\ sample.  However, when comparing $\log \lledd$ estimates from different methods in Table~\ref{tab:lledd}, we cannot exclude the possibility that the Eddington ratios adopted for  both the low- and \parent\ \imbh\ samples are systematically underestimated, by perhaps up to an order of magnitude.  If so, then nearly all of the \parent\ sample would be in the super-Eddington regime, and several of the lower-$\lledd$ objects would fall close to the proposed ``slim disk'' transition.     Furthermore, we could also be systematically underestimating $\lledd$ for our \textit{Chandra} targets if their  virial-based $\mbh$ estimates happen to be too large.    Thus, we cannot exclude the possibility that the entire \imbh\ AGN sample could feasibly be more rapidly accreting than they appear to be in Figure~\ref{fig:aox}, in which case WLQs may provide useful insight into understanding accretion onto the \imbh\ sample considered here.  We note that limited information on Eddington ratios for WLQs seems to point toward $\lledd > 0.3$; \citealt{shemmer10, luo15, plotkin15}).

Our X-ray stacking analysis of the X-ray non-detected \imbh s yields a relatively hard X-ray spectrum ($\Gamma \approx 1$). For the objects with small $\alpha_{\rm ox}$, the inner edge of the  slim disk modifies the ``intrinsic'' X-rays associated with the AGN corona, and any observed X-rays are likely dominated by reflection/scattering, such that a hard X-ray spectrum is expected.   Among WLQs, an X-ray stacking analysis of the subpopulation of X-ray weak objects reveals a hard X-ray spectrum as well \citep[$\Gamma=1.16^{+0.37}_{-0.32}$;][also see \citealt{wu12}]{luo15}.  The similarly hard (average) spectra of the subsets of X-ray weak \imbh\ AGN and X-ray weak WLQs may futher support a WLQ analogy.     Of course, the error bars on $\Gamma$ for our stacked \imbh s are large (see Table~\ref{tab:xstack}).  Still, we find this to be an intriguing result, motivating a need for tighter X-ray spectral constraints for a sample of \imbh\ AGN spanning both high- and low- $\alpha_{\rm ox}$, in order   to rigorously compare $\Gamma$ as a function of $\alpha_{\rm ox}$.

  If the WLQ analogy holds, then we require samples of \imbh s\ that are accreting even more weakly than our low-$\lledd$ sample,  in order to search for a slim-to-thin disk transition by searching for systematic changes in the X-ray properties described earlier (i.e., less dispersion in $\alpha_{\rm ox}$ and steeper $\Gamma$ at lower Eddington ratios in the thin disk regime).  The prototype \imbh\ NGC~4395 \citep{filippenko89} has a well-determined bolometric luminosity $\lbol = 5.3\times10^{40}~\ergs$ from a highly sampled broadband spectrum \citep{moran05}, providing  $\lledd \sim 0.001$  for $\mbh = \left(3.6 \pm 1.1\right) \times 10^5 \msun$ \citep{peterson05}.   Intriguingly, NGC~4395 is not only accreting below the expected slim-to-thin disk transition at 0.1--0.3 $\lledd$, but it is also near/below another critical accretion regime at $\sim$0.01$\ledd$ where the disk is expected to switch from a thin disk to a radiatively inefficient accretion flow (RIAF),\footnote{We stress that while   high-accretion rate slim disks ($\lledd \gtrsim 0.1-0.3$) and low-accretion rate RIAFs  ($\lledd \lesssim 0.01$) are both radiatively inefficient, the physical reasons for  their radiative inefficiencies are quite different.  Slim disks are radiatively inefficient largely because of photon trapping effects at near-Eddington luminosities; in most low-accretion rate RIAF models, the low radiative efficiency is mainly due to weak Coulomb coupling in the accretion flow \citep[e.g.,][]{ichimaru77, narayan94, abramowicz95}.  For other variants of RIAFs at low-accretion rates, also see, e.g..\citet{blandford99, narayan00, quataert00, merloni02}.}
  as described below.
  
 For AGN fed by a thin disk,   the X-ray photon index $\Gamma$ is correlated with Eddington ratio when $\lledd \gtrsim 0.01$ \citep[e.g.,][]{shemmer08, risaliti09, brightman13}.  Below $\sim$0.01 $\ledd$, $\Gamma$ and $\lledd$ are anti-correlated, so that AGN show the hardest X-ray spectra around $\lledd \sim 0.01$, which may indicate a transition from a thin disk to a RIAF around 1\% $\ledd$ (e.g., \citealt{constantin09, gu09, younes11, trichas13};  note that $\Gamma$ eventually plateaus to $\Gamma \sim 2.1$ at the lowest Eddington ratios, e.g., \citealt{yang15}).   This X-ray spectral behavior is observed for stellar mass black holes in X-ray binary systems as well \citep[e.g.,][]{esin97, tomsick01, wu08, sobolewska11, plotkin13, yang15}, suggesting that it is a universal feature of black hole accretion, regardless of black hole mass.   NGC~4395  supports this trend for \imbh s, as it displays a hard photon index $\Gamma= 0.61 \pm 0.15$ at $\lledd \approx 10^{-3}$ \citep{moran05}.  We have no reason to suspect that the bolometric luminosity of NGC~4395 could be biased in the same manner as for the other \imbh s (since its $\lbol$ is calculated from an observed broadband spectrum of the nucleus).  The physical mechanism for the small $\Gamma$ for NGC~4395 is therefore different than the small $\Gamma$  discussed earlier in the context of a slim disk.   This intriguing trend is of course  far from robust being based on a single source, and it further motivates a need for high signal-to-noise X-ray spectra for \imbh s across a wide range of $\lledd$.

    A sizeable population of lower-$\lledd$ \imbh\ AGN is unlikely  accessible from  optical-selection techniques, however, and recovering such objects will require complementary multiwavelength searches.   High-spatial resolution X-ray surveys (especially when combined with the radio) are  a promising avenue for revealing weakly accreting \bh s \citep[e.g.,][also note galaxies \srcfourdig\ and \srcfivedig\ in the current work that show X-ray signatures of an AGN, but lack optical photoionization signatures of activity.]{soria06, gallo10, pellegrini10, reines11, reines12, reines14, miller15, lemons15}.  Success has also been achieved through infrared surveys \citep[e.g.,][]{satyapal08}, as well as fast  variability \citep{kamizasa12, ho16, morokuma16}.

 We stress that it is unclear if the optically-based $\lledd$  measurements are  indeed systematically underestimated, and the above WLQ analogy is only meant to represent one possibility.  If the adopted  $\lledd$ estimates are  accurate, then there does not appear to be a distinct Eddington ratio that marks a transition in radiative efficiency.  In that case, a range of accretion disk/corona properties may contribute to the dispersion in $\alpha_{\rm ox}$, which may include a substantial number of intrinsically  X-ray weak AGN.  To confirm or refute current $\lledd$ estimates, broadband spectral energy distributions for a substantially larger number of \imbh s are required to properly constrain the bolometric corrections for this population.  Such a project will require high-resolution imaging across the entire electromagnetic spectrum to separate the AGN from the host galaxy (see \citetalias{dong12} for further discussion,  as well as, e.g.,  \citealt{moran99, moran05, thornton08, constantin12}, for  examples of well-sampled broadband spectra).

\section{Conclusions}
We find no evidence for a difference in the X-ray properties of \imbh s from $-2 < \log \lledd < 0$.  We argue that either  (1) optically-based $\lledd$ estimates are systematically underestimated and nearly all \imbh s accrete from a geometrically thick, radiatively inefficient slim disk; or (2) there is variety in the accretion details among individual objects, but there is no evidence for a systematic change in accretion properties at a specific Eddington ratio.   If \imbh s indeed accrete from a slim disk, then  super-Eddington accretion could provide a  mechanism for growing \smbh s  in the early Universe \citep[e.g.,][]{madau14}.  Finally,  \citet{madau15} recently showed that a faint high-redshift AGN population  could produce enough  flux to account for the epoch of reionization, provided that  AGN X-ray-to-UV luminosity ratios in the early Universe are not too hard (\citealt{madau15} explicitly adopt $\alpha_{\rm ox} \approx -1.4$).   From an empirical perspective (and regardless of the accretion mode), the  observed flattening of the $\alpha_{\rm ox}-\luv$ relation at low luminosities  implies that accretion onto \imbh s  produces a broadband continuum in line with requirements for AGN driven reioniziation.
  
\acknowledgments
We thank the referee for suggestions that improved the presentation of this paper.  We thank Ohad Shemmer for helpful discussions, and Anna Conlon for advice on our survival analysis.  Support for this work was provided by the National Aeronautics and Space Administration through Chandra Award Number GO3-14116X issued by the Chandra X-ray Observatory Center, which is operated by the Smithsonian Astrophysical Observatory for and on behalf of the National Aeronautics Space Administration under contract NAS8-03060.  Support for AER was provided by NASA through Hubble Fellowship grant HST- HF2-51347.001-A awarded by the Space Telescope Science Institute, which is operated by the Association of Universities for Research in Astronomy, Inc., for NASA, under contract NAS 5-26555.  This research has made use of software provided by the Chandra X-ray Center (CXC) in the application package CIAO.  This research has made use of data obtained from the Chandra Data Archive, and software provided by the Chandra X-Ray Center (CXC) in the application packages CIAO, ChIPS, and Sherpa.



\end{document}